\pdfoutput=1
\documentclass[format=sigconf, review=false, anonymous=false, nonacm = true, authordraft = false]{acmart}
\usepackage{enumitem}
\usepackage{tikz}
\usetikzlibrary{arrows,shapes,positioning,calc}
\usepackage{subcaption}
\usepackage{textcomp}
\usepackage[ruled,vlined]{algorithm2e}
\usepackage{tcolorbox}
\setlist[enumerate]{itemsep=0mm}
\AtBeginDocument{%
  \providecommand\BibTeX{{%
    \normalfont B\kern-0.5em{\scshape i\kern-0.25em b}\kern-0.8em\TeX}}}

\setcopyright{rightsretained}

\copyrightyear{2021}
\acmYear{2021}
\setcopyright{rightsretained}
\acmConference[FAccT '21]{Conference on Fairness, Accountability, and Transparency}{March 3--10, 2021}{Virtual Event, Canada}
\acmBooktitle{Conference on Fairness, Accountability, and Transparency (FAccT '21), March 3--10, 2021, Virtual Event, Canada}\acmDOI{10.1145/3442188.3445881}
\acmISBN{978-1-4503-8309-7/21/03}

 \usepackage{color-edits}
\addauthor[Amanda]{am}{blue}
\addauthor[Alex]{ac}{purple}
\addauthor[Dan]{dh}{green}
\addauthor[Neel]{ng}{red}
\newtheorem{definition}{Definition}
\newtheorem{assumption}{Assumption}

\newcommand{\cor}{\mathrm{cor}}
\newcommand{\cov}{\mathrm{cov}}
\newcommand{\var}{\mathrm{var}}

\newcommand{\Rn}{\mathbb{R}^n}

\begin{document}

\title{Leveraging Administrative Data for Bias Audits: Assessing Disparate Coverage with Mobility Data for COVID-19 Policy}

\author{Amanda Coston}
\email{acoston@cs.cmu.edu}
\affiliation{%
  \institution{ 
  Carnegie Mellon University}
}

\author{Neel Guha}
\email{nguha@stanford.edu}
\affiliation{%
  \institution{
  Stanford University}
}
\author{Derek Ouyang}
\email{douyang1@stanford.edu}
\affiliation{%
  \institution{
  Stanford University}
}
\author{Lisa Lu}
\email{lcl@law.stanford.edu}
\affiliation{%
  \institution{
  Stanford University}
}

\author{
Alexandra Chouldechova}
\email{achould@cmu.edu}
\affiliation{%
\institution{
  Carnegie Mellon University}}

\author{Daniel E.\ Ho}
\email{dho@law.stanford.edu}
\affiliation{%
  \institution{
  Stanford University}
}

\begin{abstract} 
    Anonymized smartphone-based mobility data has been widely adopted in devising and evaluating COVID-19 response strategies such as the targeting of public health resources.  Yet little attention has been paid to measurement validity and demographic bias, due in part to the lack of documentation about which users are represented as well as the challenge of obtaining ground truth data on unique visits and demographics.  We illustrate how linking large-scale administrative data can enable auditing mobility data for bias in the absence of demographic information and ground truth labels.
    More precisely, we show that linking voter roll data---containing individual-level voter turnout for specific voting locations along with race and age---can facilitate the construction of rigorous bias and reliability tests. 
    Using data from North Carolina's 2018 general election, these tests illuminate a sampling bias that is particularly noteworthy in the pandemic context: older and non-white voters are less likely to be captured by mobility data. We show that allocating public health resources based on such mobility data could disproportionately harm high-risk elderly and minority groups.
\end{abstract}

\settopmatter{printfolios=true}
\maketitle

\section{Introduction}
Mobility data has played a central role in the response to COVID-19.
Describing the movement of millions of people,
smartphone-based mobility data has been used to analyze the effectiveness of social distancing polices (non-pharmaceutical interventions), illustrate how movement impacts the transmission of COVID-19, and probe how different sectors of the economy have been affected by social distancing policies \citep{brzezinski2020cost, chang2020mobility, wapost_hospitals, gao2020mapping, benzell2020rationing, farboodi2020internal, killeen2020county, allcott2020polarization}.
Despite the high-stakes settings in which this data is deployed, there has been no independent assessment of the reliability of this data.
In this paper we show how administrative data 
(i.e., data from government agencies kept for administrative purposes) can be used to perform such an assessment.

Data reliability should be a foremost concern in all policy-making and policy evaluation settings, and is especially important for mobility data due to the lack of transparency surrounding data provenance.
Mobility data providers obtain their data from opt-in location-sharing mobile apps, such as navigation, weather, or social media apps, but do not disclose which specific apps feed into their data \citep{apps_know}.
This opacity prevents data consumers such as policymakers and researchers from understanding who is represented in the mobility data, a key question for enabling effective and equitable policies in high-stakes settings such as the COVID-19 pandemic.
\citeauthor{grantz2020use}
describe  ``a critical need to understand where and to what extent these biases may exist'' in their discussion on the use of mobility data for COVID-19 response.

Of particular interest is potential sampling bias with respect to important demographic variables in the context of the pandemic: age and race.
Older age has been established as an increased risk factor for COVID-19-related mortality \citep{zhou2020clinical}.
African-American, Native-American and Latinx communities have seen disproportionately high case and death counts from COVID-19 \cite{tai2020disproportionate} and the pandemic has reinforced existing health inequities that affect vulnerable communities \cite{gray2020covid}.
If certain races or age groups are not well-represented in data used to inform policy-making, we risk enacting policies that fail to help those at greatest risk and serve to further exacerbate disparities. 

In this paper we assess SafeGraph, a widely-used point-of-interest (POI)-based mobility dataset\footnote{POIs refer to anywhere people spend money or time, including schools, brick-and-mortar stores, parks, places of worship, and airports. See \url{https://www.safegraph.com/}.} for disparate coverage by age and race.
We define \textbf{coverage} with respect to a POI: coverage is
the proportion of traffic at a POI that is recorded in the mobility data.
For privacy reasons, many mobility datasets are aggregated up from the individual level to the physical POI level. 
Due to this aggregation, we lack the resolution to assess individual-level coverage quantities like the fraction of members of a demographic subgroup of interest who are represented in the data.
Nonetheless, our POI-based notion of coverage is relevant for many COVID-19 policies that are made based on traffic to POIs, such as deciding to close certain business sectors, allocating resources like pop-up testing sites to high-risk areas, and determining where to target investigations of public health order violations. We use differences in the distributions of age and race across POIs to assess demographic disparities in coverage.
 
While we focus here on a specific dataset and implications for COVID-19 policy, the question of how one can assess disparate coverage is a more general one in algorithmic governance.  Ground truth is often lacking, which is precisely why policymakers and academics have flocked toward big data, on the implicit assumption that scale can overcome more conventional questions of data reliability, sampling bias, and the like \cite{lazer2014parable, altenburger2018algorithms}. Government agencies may not always have access to protected attributes, making fairness and bias assessments challenging \cite{kallus2020assessing}.  

The main contributions of our paper are as follows:
\begin{enumerate}[itemsep=0mm]
\item We show how administrative data  can enable audits for bias and reliability (\textsection~\ref{sec:audit})
\item We characterize the measurement validity of a smartphone-based mobility dataset that is widely used for COVID-19 research, SafeGraph (\textsection~\ref{section: idealized audit}, ~\ref{sec: measurement validity results})
\item We illuminate significant demographic disparities in the coverage of SafeGraph (\textsection~\ref{sec: demographic bias})
\item We illustrate how this disparate coverage may  distort policy decisions to the detriment of vulnerable populations (\textsection~\ref{sec:policy})
\end{enumerate}


Our paper proceeds as follows. Sections~\ref{sec:related} and~\ref{sec:back} discuss related work and background on the uses of mobility data in the pandemic.  Section~\ref{sec:audit} provides an overview of our auditing framework, formalizes the assumptions to construct bias and reliability tests, and discusses the estimation approach using voter roll data from North Carolina's 2018 general election.  Section~\ref{sec:results} presents results that while SafeGraph can be used to estimate voter turnout, the mobility data systematically undersamples older individuals and minorities.  Section~\ref{sec:discuss} discusses interpretation and limitations. 

\section{Related work}
\label{sec:related}


Our assessment of disparate coverage is related to several strands in the literature.  First, the most closely related work to ours is SafeGraph's own analysis of sampling bias discussed below (\textsection~\ref{sec: safegraph analysis of sampling bias}). SafeGraph's analysis examines demographic bias only at the national aggregated level and does not address the question of demographic bias for POI-specific inferences. Ours is the first independent assessment of demographic bias to the extent we are aware.

Second, our work relates to existing work on demographic bias in smartphone-based estimates \cite{williams2015measures}. A notable line of survey research has examined the distinct demographics of smartphone users  \cite{lee2010growing, dutwin2010bias}. 
\cite{wesolowski2016connecting} and \cite{wesolowski2012heterogeneous} document significant concerns about mobility-based estimates from mobile phone data, including particularly low coverage for elderly.
The literature further finds that smartphone ownership in the United States varies significantly with demographic attributes \citep{bommakanti2020requiring}. In 2019 an estimated $81\%$ of Americans owned smartphones with ownership rates of $96\%$ for those aged 18-29 and ownership rates of $53\%$ for those aged over 65 \cite{sheet2019pew}.  Racial disparities in smartphone ownership are less pronounced, with an ownership rate of $82\%$, $80\%$, and $79\%$ for White, Latinx, and African-American individuals, respectively. Even conditional on mobile phone ownership, however, demographic disparities may still exist. App usage may differ by demographic group.  According to one report, 69\% of U.S.\ teenagers, for instance, use Snapchat, compared to 24\% of U.S.\ adults \cite{snapchat}. 
Of particular relevance to mobility datasets, the rate at which users \emph{opt in} to location sharing may vary by demographic subgroup.  \citeauthor{hoy2010gender}, for instance, reported that college-aged women exhibit greater concerns with third party data usage.  And even among users who who opt in to a specific app, usage behavior may differ according to demographics. Older users, for instance, may be more likely to use a smartphone as a ``classic phone'' \cite{andone2016age}. 

Our work responds to a recent call to characterize the biases in mobility data used for COVID-19 policies \cite{grantz2020use}. 
\citeauthor{grantz2020use} highlight the potential for demographic bias, citing ``clear sociodemographic and age biases of mobile phone ownership.'' 
They note, ``Identifying and quantifying these biases is particularly challenging, though, when there is no clear gold standard against which to validate mobile phone data.''  
We provide the first rigorous test for demographic bias using auxiliary estimates of ground truth. 

Third, our work bears similarity to the literature on demographic bias in medical data and decision-making. 
A long line of research has demonstrated that medical research is disproportionately conducted on white males \citep{dresser1992wanted, moreno2004ethnic,  shavers1997african}.
This literature has cataloged the harmful effects of making treatment decisions for subgroups that were underrepresented in the data \citep{bernal2001empirically, underwood2000minorities, vyas2020hidden}.
In much the same vein, our work calls into question research conclusions based on SafeGraph data that may not be relevant for older or minority subgroups.

Last, our work relates more broadly to the sustained efforts within machine learning to understand sources of demographic bias in algorithmic decision making \cite{barocas2017fairness, chouldechova2018frontiers, corbett2018measure, friedler2019comparative, kim2017auditing}. 
Important work has audited demographic bias of facial recognition technology \cite{buolamwini2018gender}, child welfare screening tools \cite{chouldechova2018case}, criminal risk assessment scores \cite{skeem2016risk}, and health care allocation tools \cite{obermeyer2019dissecting, altenburger2018algorithms}.  
Often the underlying data is identified as a major source of bias that propagates through the algorithm and leads to disparate impacts in the decision-making stage. 
Similarly, our study illustrates how disparate coverage in smartphone-based data can misallocate COVID-19 resources.

\section{Background on SafeGraph Mobility Data} \label{sec:back}

We now discuss the SafeGraph mobility dataset, illustrate how this data has been widely deployed to study and provide policy recommendations for the public health response to COVID-19, and discuss SafeGraph's own assessment of sampling bias.

\subsection{SafeGraph Mobility Data}

SafeGraph contains mobility data from roughly 47M mobile devices in the United States.  The company sources this data from mobile applications, such as navigation, weather, or social media apps, where users have opted in to location tracking.  It aggregates this information by points-of-interest (POIs) such as schools, restaurants, parks, airports, and brick-and-mortar stores. Hourly visit counts are available for each of over 6M POIs in their database.\footnote{See \url{https://docs.safegraph.com/docs/places-summary-statistics}.} Individual device pattern data is not distributed for researchers due to privacy concerns. Our analysis relies on SafeGraph's `research release' data which aggregates visits at the POI level.

\subsection{Use of SafeGraph Data in COVID-19 Response}

When the pandemic hit, SafeGraph released much of its data for free as part of the ``COVID-19 Data Consortium'' to enable researchers, non-profits, and governments to leverage insights from mobility data.  As a result, SafeGraph's mobility data has become the dataset de rigueur in pandemic research. 
The Centers for Disease Control and Prevention (CDC) employs SafeGraph data to examine the effectiveness of social distancing measures \citep{moreland2020timing}.  According to SafeGraph, the CDC also uses SafeGraph to identify healthcare sites that are reaching capacity limits and to tailor health communications.  The California Governor's Office, and the cities of Los Angeles \citep{city_la}, San Francisco, San Jose, San Antonio, Memphis, and Louisville, have each relied on SafeGraph data to formulate COVID-19 policy, including evaluation of transmission risk in specific areas and facilities and enforcement of social distancing measures.  Academics, too, have employed the data widely to understand the pandemic: \cite{chang2020mobility} used SafeGraph data to examine how  social distancing compliance varied by demographic group and recommend occupancy limits for business types; \cite{dave2020contagion, dave2020did} used SafeGraph to infer the effect of ``superspreader'' events such as the Sturgis Motorcycle Rally and campaign events; \cite{petrova2020divided} examined whether social distancing was more prevalent in in areas with higher xenophobia; and \cite{allcott2020polarization} examined whether social distancing compliance was driven by political partisanship, to name a few. What is common across all of these works is that they assume that SafeGraph data is representative of the target population. 

\subsection{SafeGraph Analysis of Sampling Bias}
 \label{sec: safegraph analysis of sampling bias} 
 SafeGraph has issued a public report about the representativeness of its data \citep{squire2019measuring, squire2019colab}. While SafeGraph does not have individual user attributes (e.g., race, education, income), it merged census data based on census block group (CBG) to assess bias along demographic characteristics.\footnote{CBGs are geographic regions that contain typically between 600 and 3000 residents. CBGs are the smallest geographic unit for which the census publishes data.} SafeGraph assigns each device an estimated home CBG based on where the device spends most of its nights and uses the demographics of the estimated home CBG for the bias assessment. The racial breakdown of device holders, for instance, was allocated proportionally based on the racial breakdown of the devices' estimated home CBGs.
 SafeGraph then compared the total SafeGraph imputed demographics against census population demographics at the national level. According to SafeGraph, the results suggest that their data is ``well-sampled across demographic categories'' \citep{squire2019colab}.

SafeGraph's examination for sampling bias should be applauded.  Companies may not always have the incentive to address these questions directly, and SafeGraph's analysis is transparent, with data and replication code provided.  As far as we are aware, it remains the only analysis of SafeGraph sampling bias. 

Nevertheless, their analysis suffers from several key limitations.
Most notably, this analysis does not use ground-truth demographic information and instead relies on imputed demographics using a method which suffers systematic biases. For instance, home CBG estimation is inaccurate for certain segments of the population, such as nighttime workers. 
Even when the estimated home CBG itself is correct, their imputation of \emph{demographics} from the CBG imposes a strong homogeneity assumption: The mere fact that 52\% of Atlanta's population is African American does not guarantee that five out of ten SafeGraph devices in Atlanta belong to African-Americans.

Additionally, the analysis uses an aggregation scheme which introduces two methodological limitations.
First, because their analysis aggregates CBGs nationally, the results are susceptible to undue influence from outliers, such as those resulting from errors in home CBG estimation. We anticipate these errors to be substantial since SafeGraph reports highly unrepresentative sampling rates at the CBG level, including CBGs with four times as many devices as residents.\footnote{See Fig. 3 of~\cite{squire2019colab_quantifying}.}
Second, the results may also miss significant differences in the joint distribution of features because the analysis aggregates CBGs for a single attribute at a time.
For example, if coverage is better for younger populations and for whiter populations, but whiter populations are on average older than non-white populations, then evaluating coverage marginally against either race or age will underestimate disparities. Indeed we present evidence for such an effect in \textsection~\ref{sec:results}. 

Lastly, this analysis uses CBGs as the unit of analysis which may miss disparities that exist at finer geographic units, such as POIs. This distinction is noteworthy since many of the COVID-19 analyses referenced above leverage SafeGraph data at finer geographic units than CBGs (e.g. POIs). This risks drawing conclusions from data at a level of resolution that SafeGraph has not established to be free from coverage disparities. SafeGraph warns that ``local analyses examining only a few CBGs'' should proceed with caution.
Because SafeGraph's analysis examines demographic bias only at census aggregated levels and does not address the question of demographic bias for POI-specific inferences, an independent coverage audit remains critical.
We provide such an audit using a method that uses POIs as the unit of analysis and avoids the noted methodological limitations.

\section{Auditing Framework}
\label{sec:audit}

In this section we outline our proposed auditing methodology and state the conditions under which the proposed method allows us to detect demographic disparities in coverage.  We motivate our approach by first describing the idealized audit we would perform if we had access to ground truth data. We then introduce our administrative data and subsequently modify this framework to account for the limitations of the available data.
\subsection{Notation}

Let $\mathcal{I} = \{1,...,n\}$ denote a set of SafeGraph POIs.
Let $S^j \in \Rn$ denote a vector of the SafeGraph traffic count (i.e. number of visits) for day $j \in \mathcal{J}$ where each element $S^j_i$ indicates the traffic to POI $i$ on day $j$. 
Similarly let $T^j_i$ denote the ground truth traffic (visits) to POI $i$ during day $j$.  
When the context is clear, we omit the superscript $j$ when referring to vectors $S \in \Rn$ and $T \in \Rn$.
We use $\oslash$ to denote Hadamard division (the element-wise division of two matrices).
With this, we define our coverage function $C(S,T)$.
\begin{definition}[Coverage function] \label{definition: coverage function}
Let $C(S,T) \colon \Rn \times \Rn \mapsto \Rn$ denote the following coverage function: 
\begin{align*}
    C(S,T) = S \oslash T 
\end{align*}
The coverage function yields a vector where the ith element equals $\frac{S_i}{T_i}$ and describes the coverage of POI i.
\end{definition}
Let $D^j_i$ denote a numeric measure of the demographics of visitors to POI $i$ on day $j$; for instance $D^j_i$ may be the percentage of visitors to a location on a specific day that are over the age of 65. 
Let $\cor(X,Y) = \frac{\cov(X,Y)}{\sqrt{\var(X)\var(y)}}$ denote the Pearson correlation between vectors $X$ and $Y$ and let $r(X)$ be the rank function that returns the rank of vector $X$.\footnote{The rank assigns each element of the vector the value of its rank in an increasing ranking of all elements in the vector. For example the rank of vector "(5, 1, 3)" would be "(3, 1, 2)".}
Our audit will consider the (Spearman) rank correlation $\cor(r(X), r(Y))$, which provides a more flexible and robust measure of association than the Pearson correlation.
\subsection{Idealized Audit} \label{section: idealized audit}
Our audit assesses how well SafeGraph measures ground truth visits and whether this coverage varies with demographics. We operationalize these two targets as follows:
\begin{definition}[Measurement signal and validity] \label{definition: measurement validity}
Define the \textbf{strength of measurement signal} as 
\[
\cor(r(S), r(T)).
\]
\end{definition}
A positive signal indicates facial measurement validity, and a signal close to one indicates high measurement validity.

\begin{definition}[Disparate coverage] \label{definition: disparate coverage}
We will say that \textbf{disparate coverage} exists when the rank correlation between coverage and the demographic measure is statistically different from zero: 
\begin{align*}
  \cor \big( r(C(S, T), r(D) \big) \neq 0.
\end{align*}
\end{definition}

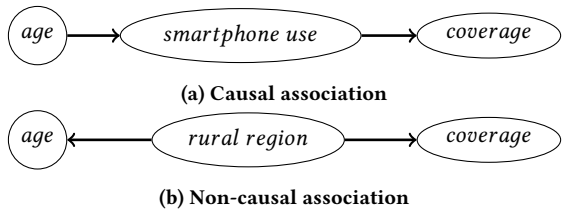
\begin{figure}[t]
    \centering
    \begin{subfigure}[b]{0.95\columnwidth}
      \centering
\begin{tikzpicture}
\node[circle,draw=black] (age) at (0, 2) {$age$};
\node[ellipse,draw=black] (cell) at (2.7,2) {$smartphone\ use$};
\node[ellipse, draw=black] (cov) at (6, 2) {$coverage$};

\draw[->,line width= 1] (age) --(cell);
\draw[->,line width= 1] (cell) -- (cov);
\end{tikzpicture}
\caption{Causal association}
\end{subfigure}

\begin{subfigure}[b]{0.95\columnwidth}
      \centering
      \begin{tikzpicture}
\node[circle,draw=black] (age) at (0, 0) {$age$};
\node[ellipse,draw=black] (rural) at (2.8,0) {$rural\ region$};
\node[ellipse, draw=black] (cov) at (6, 0) {$coverage$};

\draw[->,line width= 1] (rural) --(age);
\draw[->,line width= 1] (rural) -- (cov);
\end{tikzpicture}
\caption{Non-causal association}
\end{subfigure}
\caption{Possible mechanisms under which disparate coverage arises. Disparate coverage may be a result of a causal associations such as (a) whereby older people are less likely to own or use smartphones and therefore places frequented by older people have lower coverage. Disparate coverage may also arise due to a non-causal associations such as (b) whereby rural regions have higher percentages of older residents and worse cell reception which reduces coverage. Both types of associations are policy-relevant because in both cases, certain age groups are underrepresented.} 
    \label{fig:associations}
\end{figure}

We are interested in identifying an association of any kind; we are not concerned with identifying a causal effect per se. 
Age might have a causal effect on smartphone usage, setting aside the question of manipulability \cite{holland1986statistics}, as depicted in the top panel (a) of Fig.~\ref{fig:associations}.  But as the bottom panel (b) depicts, age may not directly affect SafeGraph coverage but be directly correlated with a factor like urban/rural residence, which in turn does affect SafeGraph coverage. For either mechanism, the policy-relevant conclusion remains that SafeGraph is underrepresenting certain age groups.  

In reality, there is no ground truth source of information about foot traffic and the corresponding demographics for all 6 million POIs.
Instead, we must make do with estimates of $T$ and $D$ based on auxiliary data sources about some subset of visits 
to a subset of POIs.
In order to identify the relationship of interest (Def.~\ref{definition: disparate coverage}) between coverage and demographics, we need the following to hold:\footnote{Appendix~\ref{app: assumptions required for measurement validity} discusses the analogous assumptions required to identify the target for measurement validity (Def.~\ref{definition: measurement validity}).}
\begin{definition}[No induced confounding] \label{assumption:confounding}
  The estimation procedure does not induce a confounding factor that affects both the estimate of demographics and the estimate of coverage.
\end{definition}
\begin{definition}[No selection bias]
\label{assumption:selection}
  The selection is not based on an interaction between factors that affect coverage and demographics.
\end{definition}

We emphasize the difficulty in obtaining this information.
It is challenging to obtain estimates of foot traffic to POIs. In fact, researchers typically treat smartphone-based mobility data as if it were ground truth (e.g. \cite{bao2020covid}). It is even more challenging to identify data sources for ground truth visits to POIs \emph{with} corresponding demographic information \cite{grantz2020use}. 
Consider for instance large sporting events where stadium attendance is closely tracked. 
Can we leverage differences in audience demographics based on the event (e.g., international soccer game between two countries) in order to assess disparate coverage?
Two major impediments are lack of access to precise demographic estimates as well as confounding factors such as tailgating that may vary with demographics.

\subsection{Administrative data on voter turnout}

We propose a solution using large-scale administrative data that records individual-level visits along with demographic information: voter turnout data in North Carolina's 2018 general election from L2, a private voter file vendor which aggregates publicly available voter records from jurisdictions nationwide.\footnote{See \url{https://l2political.com/}.}  
Our analysis relies primarily on four fields in the L2 voter files: age, race, precinct, and turnout. The L2 data is missing one key piece of information: the poll location. We use a crosswalk of voting precinct to poll location obtained from the North Carolina Secretary of State to map each voter via their voting precinct to a SafeGraph POI.
 Overall, our data includes 539K voters who turned out to vote at 558 voting locations that could be matched. Table~\ref{table:voter_stats} presents summary statistics on voters associated with polling locations that could be matched, showing that our data is highly representative of all voting locations. (Details on the data and preprocessing are provided in Appendices~\ref{app:data} and~\ref{app:preprocess}.)

\begin{table}[ht]
\centering
\begin{tabular}{lrr}
  \hline
 & Matched Voters & All Voters \\ 
  \hline
Voters & 539,607 & 1,581,937 \\ 
  Mean Age & 52.57 & 52.78 \\ 
   Std Age & 16.67 & 16.59 \\ 
  Proportion over 65 & 0.25 & 0.26 \\ 
  Proportion Hispanic & 0.04 & 0.04 \\ 
  Proportion Black & 0.20 & 0.19 \\ 
  Proportion White & 0.70 & 0.71 \\ 
   \hline
\end{tabular}
\caption{Demographics of all voters in North Carolina's 2018 general election compared to voters included in our analysis ("matched voters").
The matched voters are representative of the full voting population. Details of the matching procedure are given in Appendix~\ref{app:preprocess}.} 
\label{table:voter_stats}
\end{table}

Derived from official certified records by election authorities, voter turnout information is of uniquely high fidelity. In an analysis of five voter file vendors, Pew Research, for instance, found that the vendors had 85\% agreement about turnout in the 2018 election \citep{igielnik2018commercial}.  Voter registration forms typically include fields for date of birth, gender, and often race.\footnote{North Carolina, for instance requests both race and ethnicity (\url{https://s3.amazonaws.com/dl.ncsbe.gov/Voter_Registration/NCVoterRegForm_06W.pdf}).} When race is not provided, data vendors estimate race. The Pew study found race to be 79\% accurate across the five vendors, with accuracies varying from 67\% for African-Americans to 72\% for Hispanics to 93\% for non-Hispanics.\footnote{The study did not name which voter file vendors were analyzed.} 
We can identify individuals visiting a specific voting location on election day because North Carolina differentiates \emph{in person, election day} voters from absentee, mail, and early voters. 
We note that poll locations are often schools, community centers, religious institutions, and fire stations. 
These POIs may hence also have non-voter traffic on election day.
We address this possible source of confounding by adjusting the SafeGraph traffic using an estimate of non-voter traffic.

\subsection{Adjustment for non-voter traffic}
\label{sec:conf}

Non-voter traffic may be incorporated into SafeGraph measures and may confound our analysis if the magnitude of that non-voter traffic varies with the demographic attributes of the voters.
For instance, if younger voting populations are more likely to vote at community centers which have large non-voter traffic and older voting populations are more likely to vote at fire stations which have small non-voter traffic, then even if SafeGraph has no disparate coverage, we would observe a negative relationship between coverage and age.\footnote{Non-voter traffic may be affected by device attribution errors, in which device GPS locations are incorrectly assigned to one of two adjacent POIs. SafeGraph reports in its user documentation that "[it] is more difficult to measure visits to a midtown Manhattan Starbucks than a visit to a suburban standalone Starbucks." If younger voting populations are more likely to vote in dense urban polling locations, then even if there isn't large non-voter traffic in the same facility, large traffic in an adjacent facility could still be incorrectly attributed to the polling location with greater likelihood than to a suburban polling location. However, this source of confounding can be controlled for using the same technique described.}
We control for this confounding by estimating non-voter traffic using mean imputation.
In Appendix~\ref{app:bandwidth}, we provide similar results using a linear regression imputation procedure. 

\subsubsection{Additional notation}
Letting $j^*$ denote election day,
we estimate the non-voter traffic at poll location $i$ on election day, $Z_i^{j^*}$, by averaging SafeGraph traffic to $i$ on adjacent days: $${Z_i^{j^*} = { \frac{S_i^{j^* -1} + S_i^{j^* + 1}}{2}}}$$ 
This adjustment enables us to compute the marginal traffic over the estimated baseline, which we term SafeGraph marginal traffic.\footnote{The adjustment resulted in negative estimates of voter traffic for poll locations at schools. In the Appendix~\ref{app:preprocess}, we show that baseline traffic estimation is generally much worse for school, due in part to school holidays or large-scale events such as sports games. As a result, we exclude schools from our analysis.}
\begin{definition}[Marginal traffic]
SafeGraph marginal traffic denotes device counts above estimated baseline: $S_i^{j^*} - Z_i^{j^*}$. 
\end{definition}
Let $V^{j*}_i$ denote the number of voters at poll location $i$ as recorded by L2.
With this, we refine our definition of coverage using the coverage function from Def.~\ref{definition: coverage function}: 
\begin{definition}[SafeGraph coverage]
SafeGraph coverage is $C(S^{j^*} - Z^{j^*}, V^{j^{*}})$.
Each element $i$ of this vector refers to the ratio of marginal traffic at POI $i$  to voter turnout at $i$. 
\end{definition}

\subsection{Audit via voter turnout}
\label{sec:test_voters}
 The disparate impact question in this setting is \emph{does SafeGraph coverage of voters at different poll locations vary with voter demographics?}
We focus on two key demographic risk factors for COVID-19: age and race. 
We summarize the age distribution at a polling location $i$ by computing the proportion of voters over age 65. 
For race, we consider the proportion of voters who are an ethnic group besides white.\footnote{In what follows we use the generic variable $D$ to indicate either measure of demographics.}

Def.~\ref{definition: disparate coverage} formalizes this question as testing 
whether there is a rank correlation between ${C(S^{j*} - Z^{j*}, V^{j*})}$ and demographic measure $D$. However such a test may be misleading if we have induced confounding by our estimation procedure (Def.~\ref{assumption:confounding}). We can incorporate a test of confounding into our audit. Specifically, we can test for \emph{time-invariant} confounding. 
\begin{definition}[time-invariant confounding]
\label{definition: time-invariant confounding}
A time-invariant confounder affects our demographic estimate as well as traffic on election day \emph{and} on non-election days.
\end{definition}
This contrasts to a \textbf{time-varying confounding}:
\begin{definition}[time-varying confounding]
\label{definition: time-varying confounding}
A time-varying confounder affects our demographic estimate and traffic on election day only. It does \emph{not} affect traffic on non-election days.
\end{definition}
Examples of time-invariant and time-varying confounding are given in Figure~\ref{fig:time_conf}. 
The assumption of no time-varying confounding is untestable but it is reasonable to believe this holds in our setting. 
Most voting places, for instance, are public places making it unlikely that the non-voter traffic is affected differentially on election and non-election days. 
Another possible time-varying confounder would be if voting locations with older (or largely non-white) voters are more likely to be placed outside of the SafeGraph geometry for device attribution (e.g., parking lot).  
We do not believe this is likely because voting locations are typically indoors for security and climate reasons during a November election.
We can accommodate time-invariant confounding in our audit by modifying the definition of disparate coverage.

\begin{figure*}[t]
    \centering
       \begin{subfigure}[b]{\columnwidth}
       \centering
\begin{tcolorbox}[colback=blue!5!white,colframe=blue!10!gray,  colbacktitle=blue!60!white, title=Time-invariant confounding]
 \textbf{Example:} Younger voting populations vote at places like community centers with large non-voter traffic whereas older populations vote at places like fire stations with little non-voter traffic. \phantom{on election day}
 \tcblower
 Testable (see \textsection~\ref{sec:test_voters})
\end{tcolorbox}
     \end{subfigure}
     \begin{subfigure}[b]{\columnwidth}
     \centering
     \begin{tcolorbox}[colback=blue!5!white,colframe=blue!10!gray,
  colbacktitle=blue!60!white,title=Time-varying confounding]
 \textbf{Example:} Younger voting populations vote at places that are open to non-voter traffic on election day whereas older populations vote at places that are closed to non-voter traffic on election day
  \tcblower
  Untestable assumption
\end{tcolorbox}
  \end{subfigure}
   \caption{We distinguish between two types of confounding: time-invariant versus time-varying confounding. We test for time-invariant confounding  (\textsection~\ref{sec:test_voters}) but we cannot test for time-varying confounding. Our results assume no time-varying confounding.}
    \label{fig:time_conf}
\end{figure*}

\begin{definition}[Disparate coverage] \label{definition: robust disparate coverage}
We will say that \textbf{disparate coverage} exists when the rank correlation between coverage on \textbf{election} day and voter demographics is statistically different from the rank correlation between coverage on \textbf{non-election} day and voter demographics: For $j \neq j*$,
\begin{align*}
  &\cor \big(r({C(S^{j*} - Z^{j*}, V^{j*})}, r(D^{j*}) \big) \neq \\
  &\cor \big(r({C(S^{j} - Z^{j}, V^{j*})}, r(D^{j*}) \big)
\end{align*}
\end{definition}

We evaluate this more robust notion of disparate coverage 
using 40 weekdays in October and November of 2018 to generate a placebo distribution of the estimated correlation coefficients against which we compare the election-day estimate.\footnote{We use weekdays in October and November except October 1 and November 5, 7, and 30. November 5 and 7 are adjacent to election day, so the baseline adjustment (\textsection~\ref{sec:conf}) would be biased. Out of convenience, we drop October 1 and November 30 to avoid having to pull September and December data to respectively compute their baseline adjustments.}
Algorithm~\ref{alg:placebo inference disparate coverage} provides details (note that
$\mathbb{I}$ denotes the indicator function). This procedure is similar to methods of randomization inference in the literature on treatment effects \cite{ho2006randomization}.
If the election-day correlation is unlikely under placebo distribution (i.e. small $p$-value), and we additionally believe there is no time-varying confounding, then we can conclude that SafeGraph has disparate coverage of voters on election day.

\begin{algorithm}[htbp!]
\SetAlgoLined
\KwIn{Voter data $(V^{j*}, D^{j*})$
SafeGraph data $\{(S^{j}, Z^{j})\}_{j=1}^{n}$
}
\KwResult{$p$-value for the election-day correlation under the placebo distribution}
\For{$j = 1,2, \hdots n$}{ 
Compute $\rho_j = \cor(r(C(S^{j} - Z^{j}, V^{j*})), r(D^{j*}))$.
}
\Return{$p = \frac{1}{n} \displaystyle \sum_{j =1}^n\mathbb{I}\{(\rho_j \leq \rho_{j^{*}})\}$}
\caption{Assessing Disparate Coverage (Def.~\ref{definition: robust disparate coverage})}
\label{alg:placebo inference disparate coverage}
\end{algorithm}

In order to generalize these findings to the broader population on non-election day, the selection cannot be based on factors that affect both coverage and demographics (See Def.~\ref{assumption:selection}).
Example violations might include: (i) The older (or non-white) population that doesn't vote is more likely to use smartphones than the older (or non-white) population that does vote; and (ii) Older (or non-white) voters leave their smartphones at home when they go vote but always carry their smartphones otherwise, whereas younger (or white) voters bring their smartphones to the polls and elsewhere.  We believe such mechanisms are unlikely.
Testing this assumption would require the use of an additional auxiliary dataset which is outside the scope of this paper. We emphasize that this assumption of no selection bias can still hold even though the voting population is not a random sample of the population with respect to demographics.
In fact, since the voting population is older and more white than the general population \cite{igielnik2018commercial}, the association between coverage and age/race among voters could very well underestimate the magnitude of the population association.

We should also consider the association between demographics age and race. It is well known that younger populations have a larger proportion non-white relative to older populations, and this holds in our sample. Polling locations with younger voters are also more likely to have higher proportions of minority voters (Appendix~\ref{app:data}). Additionally, there are widespread concerns that disparate impact can be more pronounced at the ``intersection'' of protected groups \cite{crenshaw1989demarginalizing, cabrera2019fairvis, buolamwini2018gender}. We can jointly test for disparate coverage by modifying Alg.~\ref{alg:placebo inference disparate coverage} for the multiple regression setting. We perform $n = 41$ linear regressions to model coverage as a function of the percentage over 65 and the percentage non-white
for each weekday $j$ in October and November 2018. 
We test whether the election-day coefficients on age/race are different from the $40$ non-election day coefficients on age/race. Alg.~\ref{alg:joint placebo inference disparate coverage} provides details using the notation that $A^{j*}$ denotes the proportion of voters over age 65 and $R^{j*}$ denotes the proportion of voters who are non-white.  
Code is available at \href{https://github.com/mandycoston/covid-mobility-disparate-coverage}{https://github.com/mandycoston/covid-mobility-disparate-coverage}. 

\begin{algorithm}[htbp!]
\SetAlgoLined
\KwIn{Voter data $(V^{j*}, D^{j*})$
SafeGraph data $\{(S^{j}, Z^{j})\}_{j=1}^{n}$
}
\KwResult{$p$-values for the election-day coefficients on race and age under the placebo distribution}
\For{$j = 1,2, \hdots n$}{ 
Fit a linear regression: $C(S^{j} - Z^{j},V^{j*}) =  \alpha_j A^{j*},  + \beta_j R^{j*} + \gamma_j$
}
\Return{$p_A = \frac{1}{n} \displaystyle \sum_{j =1}^n\mathbb{I}\{(\alpha_j \leq \alpha_{j^{*}})\}$ and $p_R = \frac{1}{n} \displaystyle \sum_{j =1}^n\mathbb{I}\{(\beta_j \leq \beta_{j^{*}})\}$} 
\caption{Assessing Joint Disparate Coverage}
\label{alg:joint placebo inference disparate coverage}
\end{algorithm}

\section{Results} 
\label{sec:results}

\subsection{Measurement Validity} \label{sec: measurement validity results}
\begin{figure}[tb]
    \centering
    \includegraphics[scale = 0.4]{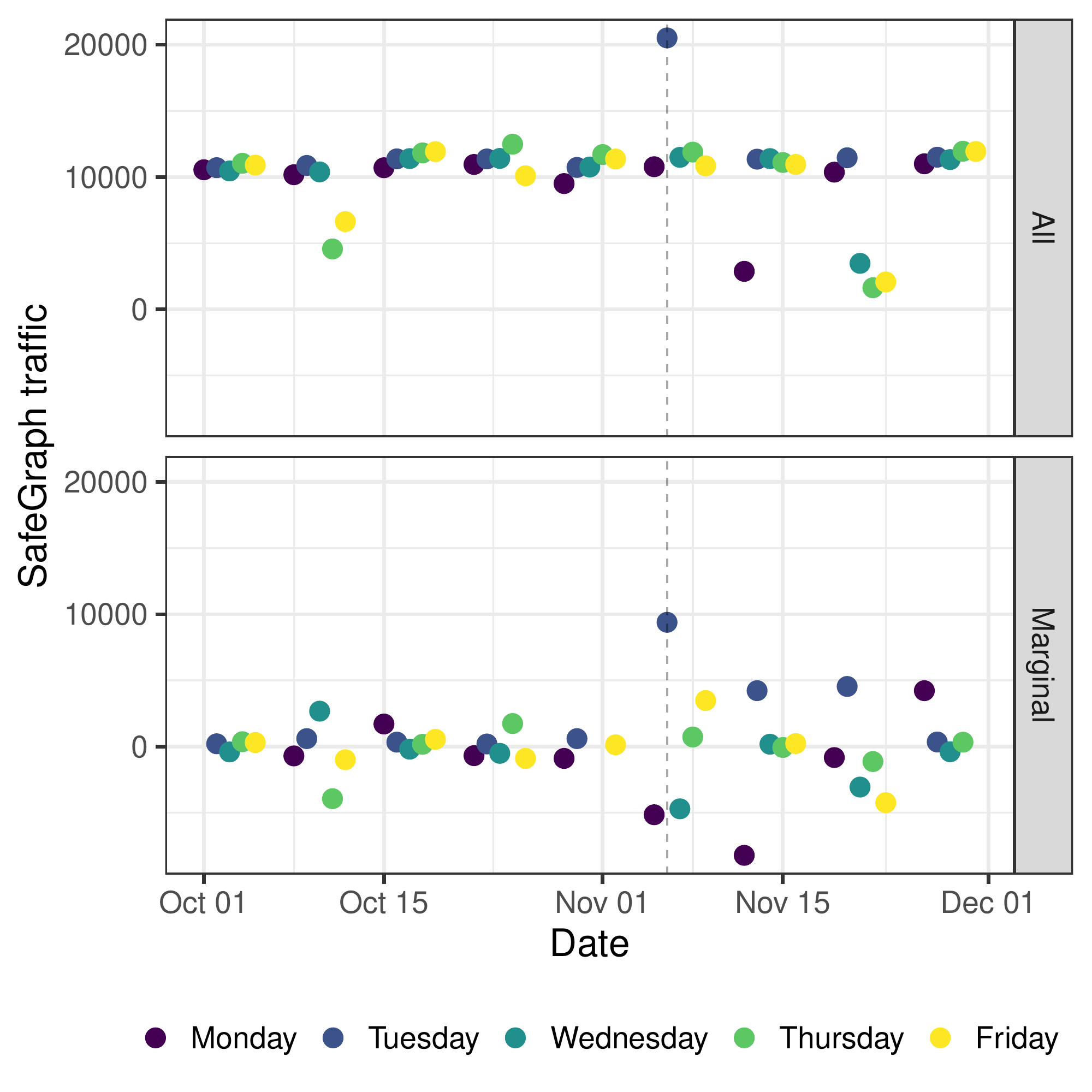}
    \caption{SafeGraph traffic by weekday over October and November 2018 for all polling locations in North Carolina. The top panel shows all SafeGraph traffic and the bottom panel shows the marginal traffic computed using the method in \textsection~\ref{sec:conf}. In both total and marginal traffic, the election day (dotted) line shows a significant boost in traffic.}
    \label{fig:monthly_traffic}
\end{figure}

\begin{figure}[tb]
    \centering
    \includegraphics[scale = 0.5]{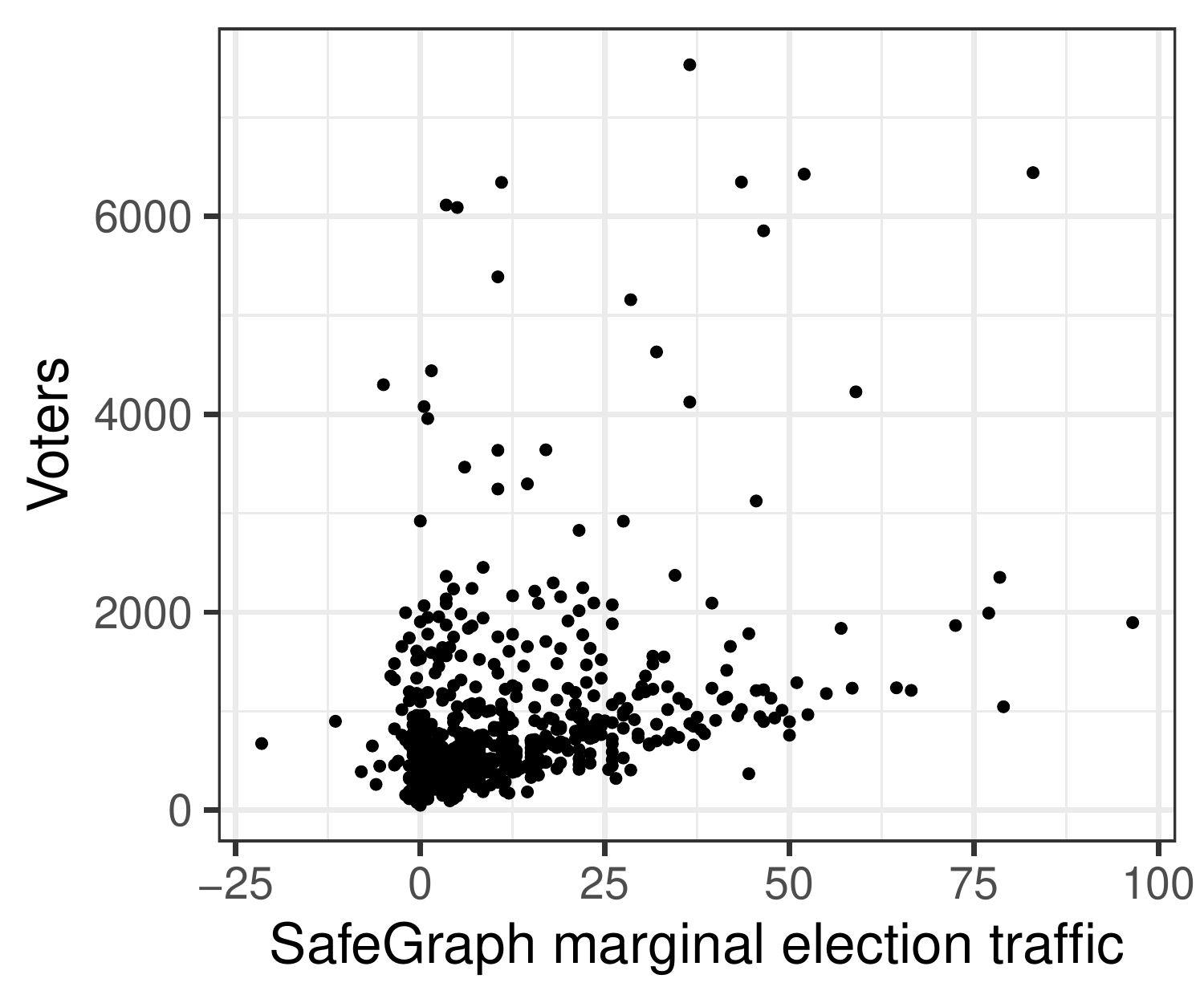}
    \caption{Election day traffic as observed by SafeGraph ($x$-axis) and actual voter turnout across polling locations ($y$-axis). Each dot represents a polling location in North Carolina in the 2018 general election.}
    \label{fig:sg_vs_voters}
\end{figure}

Election day brings a dramatic increase in traffic to polling locations relative to non-election days, and any valid measure of visits should detect this outlier.
Figure~\ref{fig:monthly_traffic} shows the daily aggregate traffic across poll locations for October and November of 2018, and as expected, we see a significant increase in both total traffic (top panel) and marginal traffic (bottom panel) on election day. 
To assess the strength of this signal using the framework described above (Def.~\ref{definition: measurement validity}), we present the correlation between marginal SafeGraph traffic on election day and actual voter turnout.
The rank correlation test yields a positive correlation: $\cor\big(r(S^{j*} - Z^{j*}), r(V^{j*})\big) = 0.383$ with $p$-value $ < 0.001$.\footnote{The correlation is similar but slightly lower for unadjusted SafeGraph traffic: $\cor\big(r(S^{j*}), r(V^{j*})\big) = 0.373$ with $p$-value $ < 0.001$.}  
Figure~\ref{fig:sg_vs_voters} displays this relationship by comparing the marginal election traffic $S_i^{j^*} - Z_i^{j^*}$ on the $x$-axis against actual voter counts $T_i^{j^*}$ on the $y$-axis for each polling location.

This corroborates that SafeGraph data is able to detect broad patterns in movement and visits.  That said, the estimates at the individual polling place location level are quite noisy: root mean-squared error is 1375 voters. 
For instance, amongst polling places that registered 20 marginal devices, roughly 300 to 2300 actual voters turned out.
This significant noise is likely due to a combination of factors.
First, SafeGraph may incorrectly attribute voters to nearby POIs because of incorrect building geometries.
Second, we may not be able to perfectly adjust for non-election traffic.
Third, SafeGraph may have disparate coverage of voters by demographic attributes.
This last factor is the focus of our analysis.

\subsection{Demographic Bias}
\label{sec: demographic bias}
\begin{figure}[tb]
    \centering
    \includegraphics[scale=.3]{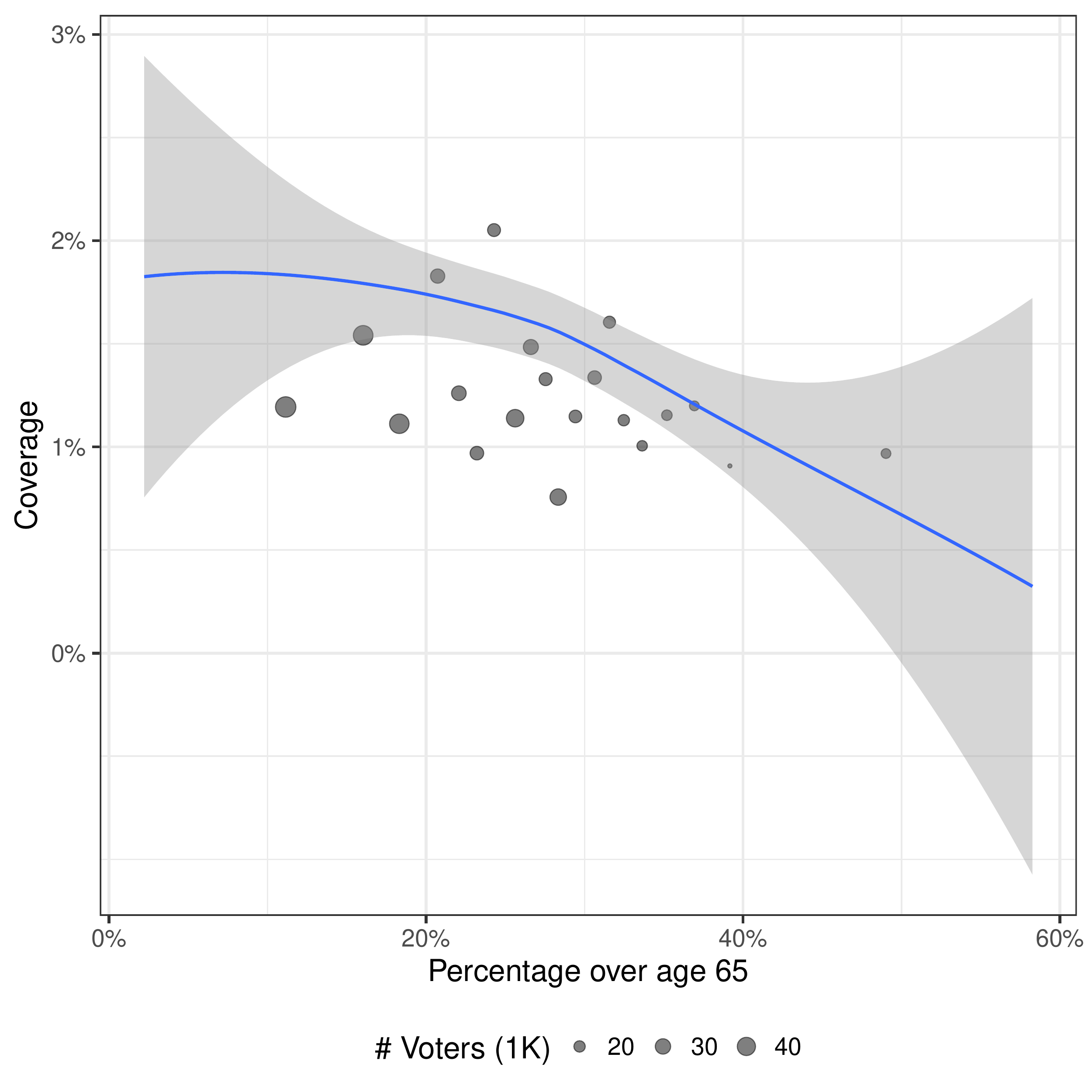}
    \includegraphics[scale=0.3]{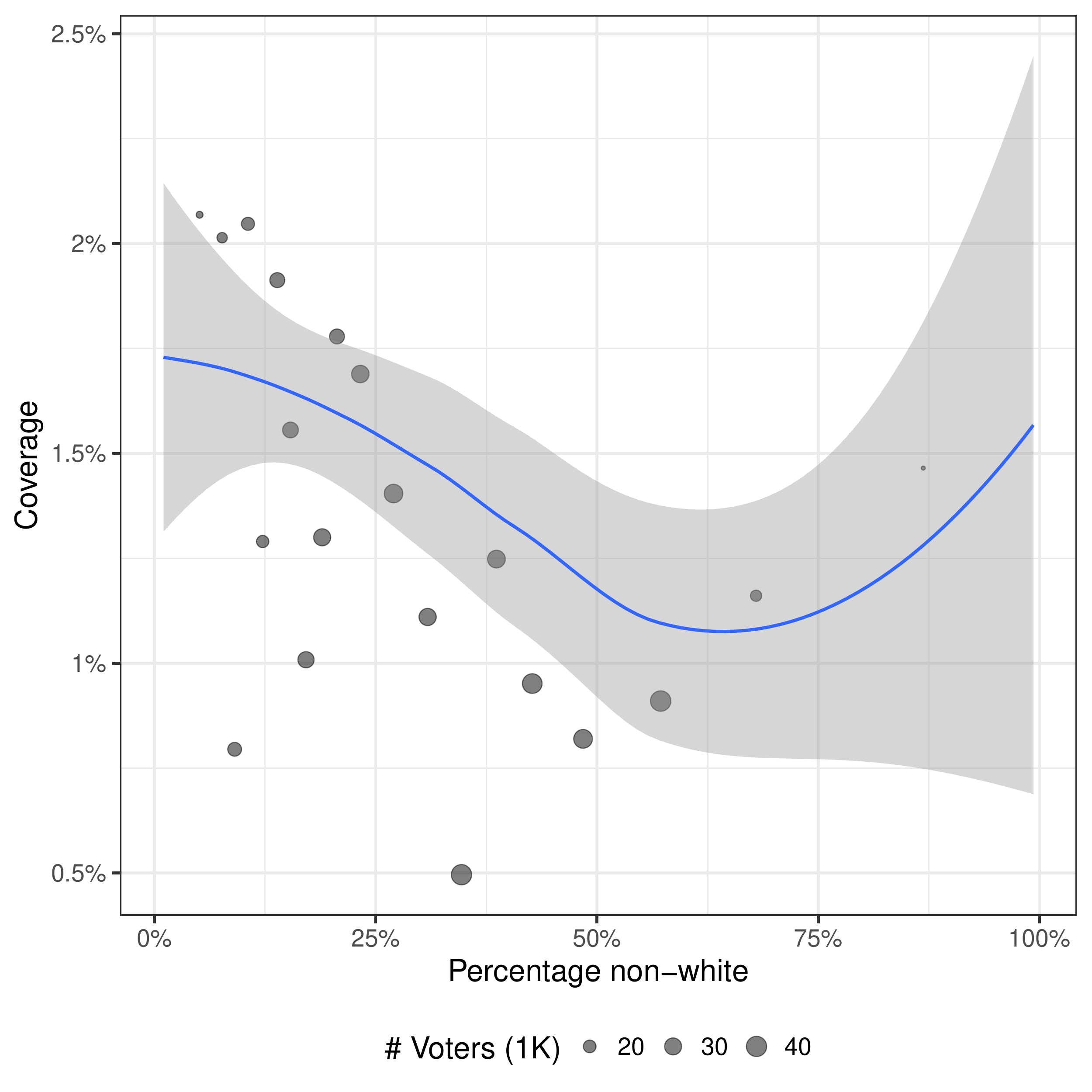}
    \caption{Estimated SafeGraph coverage rates against age and race for North Carolina 2018 general election. Each point displays a ventile of poll location by age (top) and race (bottom). The blue lines depict LOESS smoothing on the individual poll locations.}
    \label{fig:age_rate}
\end{figure}

We assess whether the demographic composition of voters who actually turned out to vote in person is correlated with coverage.  
We begin with preliminary results and then proceed to our main disparate coverage results (as defined in Def.~\ref{definition: robust disparate coverage}). 

\subsubsection{Preliminary results}
\begin{figure}[t]
    \centering
    \includegraphics[scale=0.3]{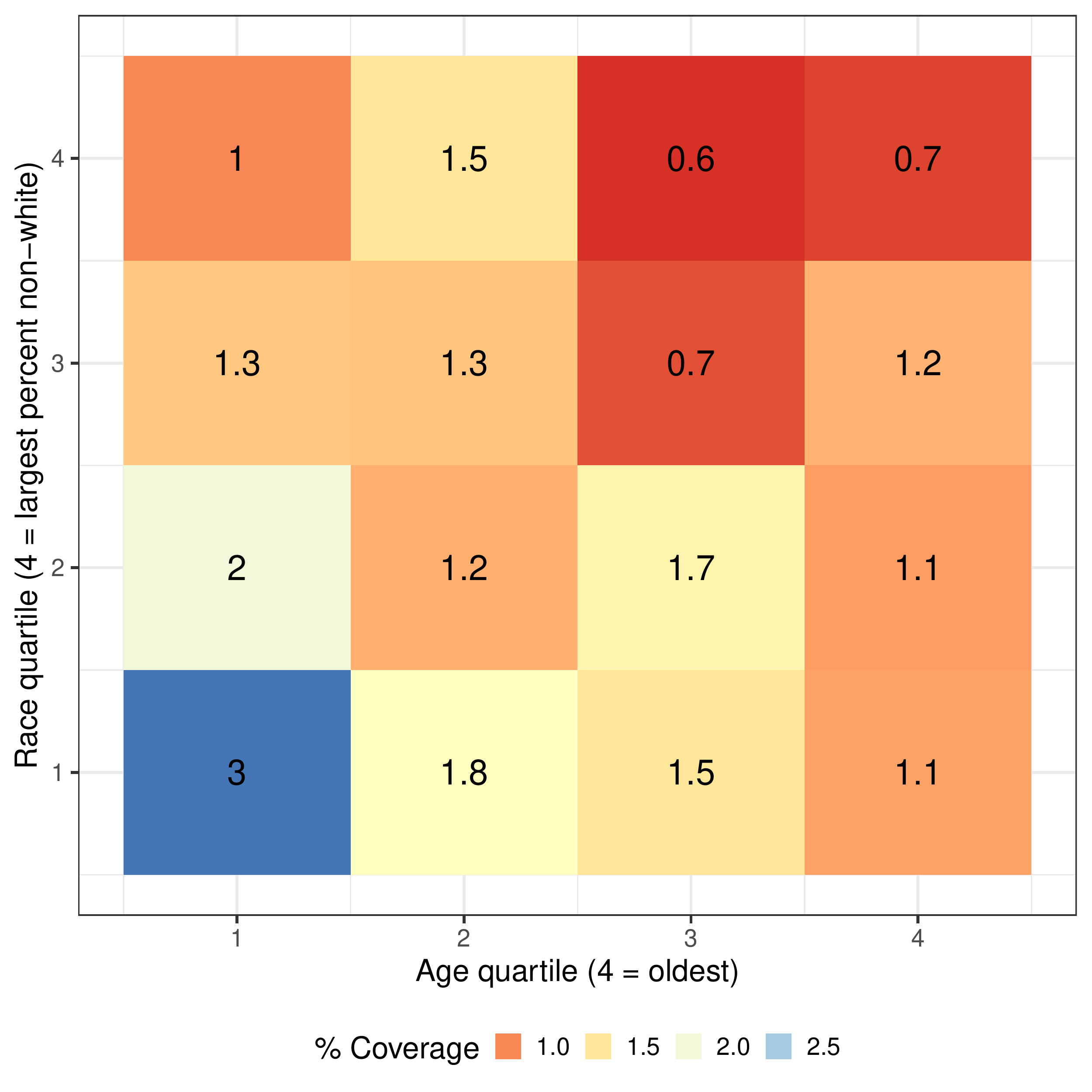}
    \includegraphics[scale= 0.31]{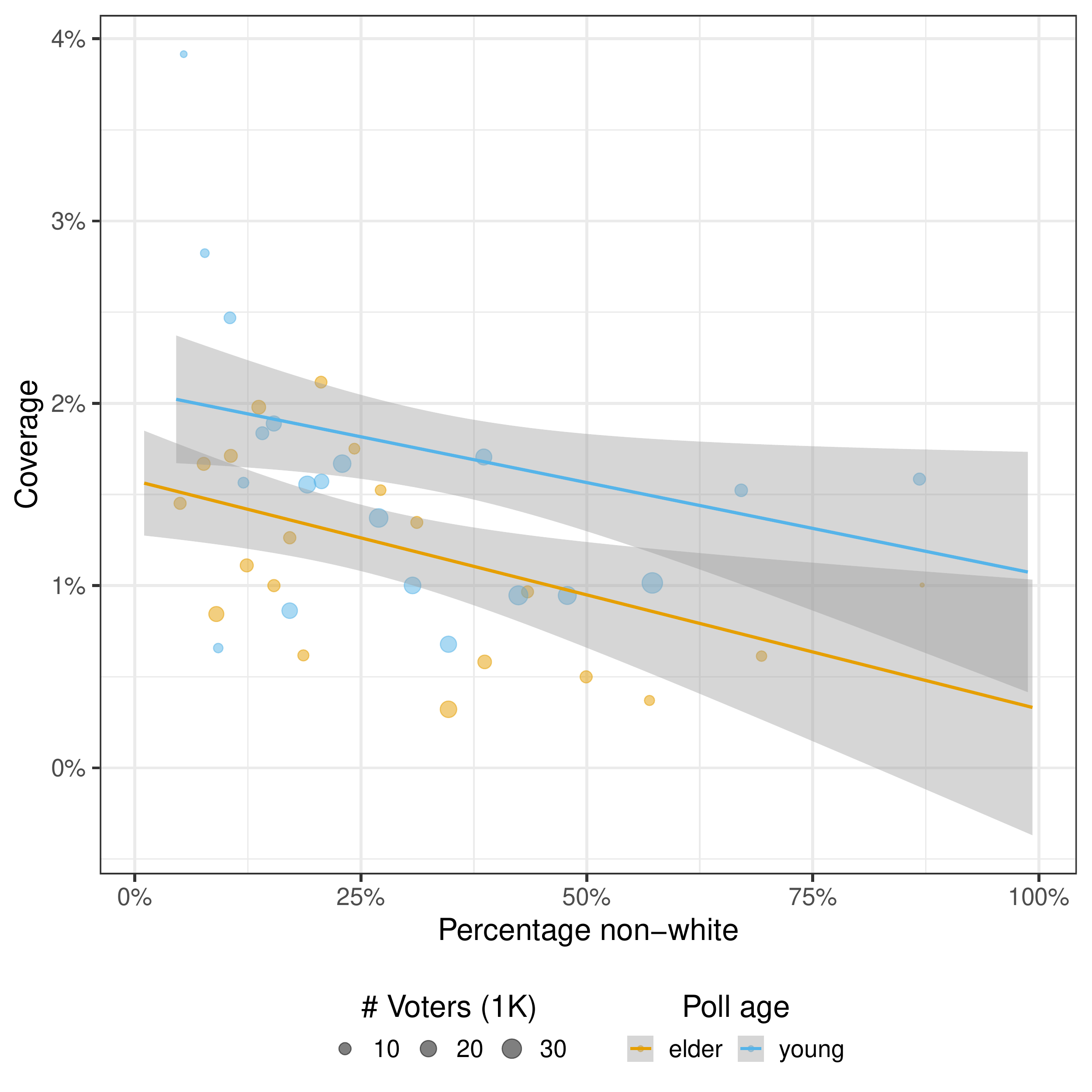}
    \caption{Intersectional coverage effects by race and age. The top panel presents the coverage rate by quartiles of age on the $x$-axis and race on the $y$-axis.  The bottom panel plots the coverage rate on the $y$-axis against percentage of non-white voters at the polling location on the $x$-axis for older polling locations (yellow) versus younger polling locations (blue) for ventiles of poll location by race. (Lines display linear smoothing of the individual poll locations.) Coverage is lowest among older minority populations and highest among younger whiter populations. }
    \label{fig:heat_map}
\end{figure}

Polling locations with older votes have lower coverage rates. 
The top panel of Figure~\ref{fig:age_rate} shows how SafeGraph coverage $C(S-Z,V)$ varies with $A$, the proportion of voters over age 65. 
The rank correlation test yields $\cor\big( r(C(S-Z,V), r(A) \big) = -0.14$ with $p$-value $<0.001$. 
We also show how coverage decreases as the proportion of non-white voters increases (bottom panel).
The rank correlation of race and coverage is $\cor\big( r(C(S-Z,V), r(R) \big) = -0.11$ with $p$-value $=0.0067$.  
 The top panel of Figure~\ref{fig:heat_map} presents a heat map of coverage with age bins (quartiles) on the $x$-axis and race bins (quartiles) on the $y$-axis.  
This bottom left cell, for instance, shows that precincts that are the most white and young have highest coverage rates.  
The lowest coverage is for older minority precincts.  
The lower panel of Figure~\ref{fig:heat_map} similarly plots race on the $x$-axis against coverage on the $y$-axis, separating older precincts (yellow) and younger precincts (blue). Older precincts on average have lower coverage rates than younger precincts, and coverage declines as the minority population increases. 

\subsubsection{Main results} \label{sec:robust test results}
\begin{figure}
    \centering
    \includegraphics[scale=0.3]{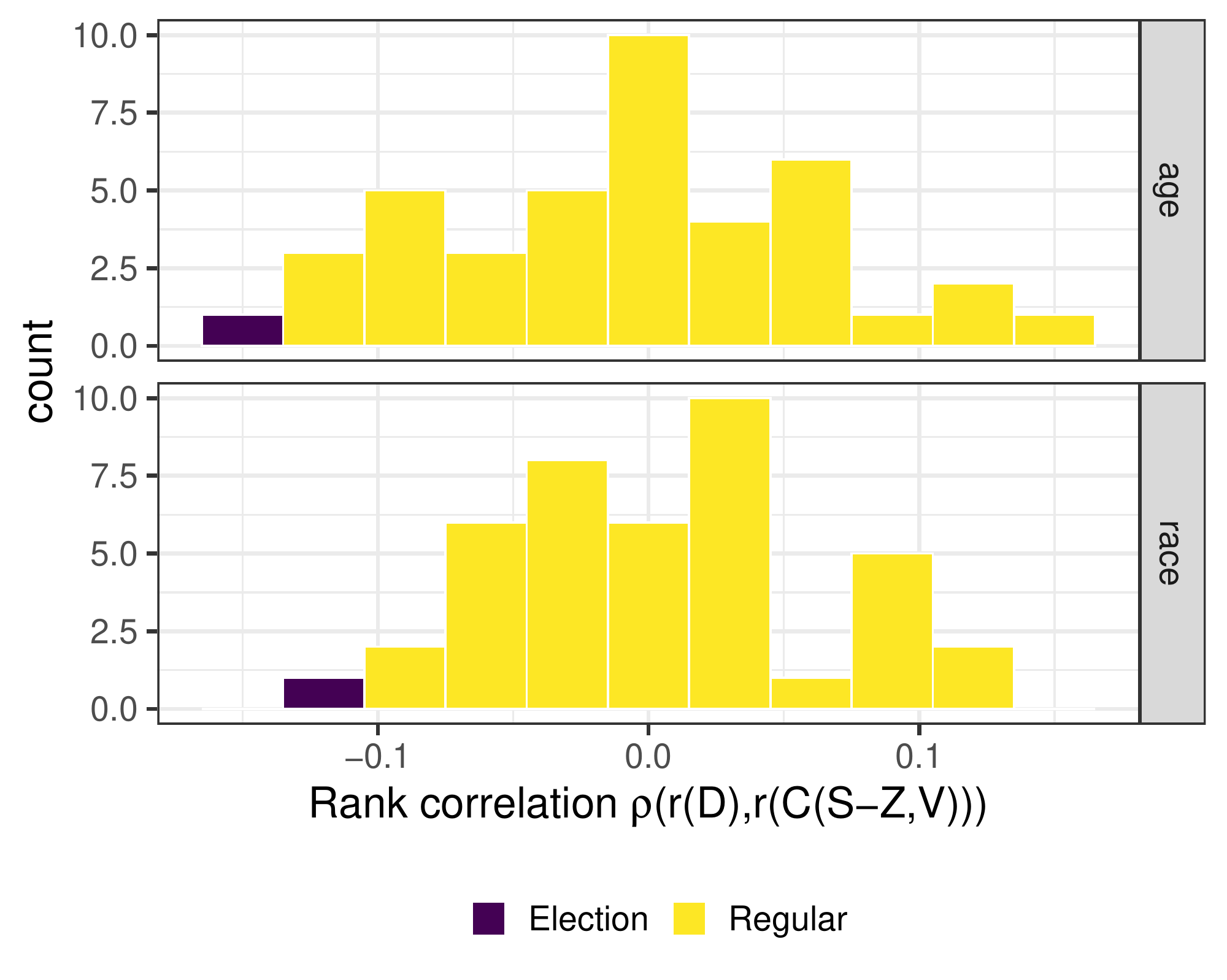}
    \caption{Distribution of placebo rank correlations between election-day demographics and marginal SafeGraph traffic on non-election days.  
    Under the empirical placebo distribution, the election-day coverage's negative correlations with age (top panel) and race (bottom panel) are very unlikely  ($p$-value $< 0.05$). Placebo correlations computed for 40 weekdays in October and November 2018.}
    \label{fig:placebo_cor}
\end{figure}

\begin{figure}
    \centering
    \includegraphics[scale =0.3]{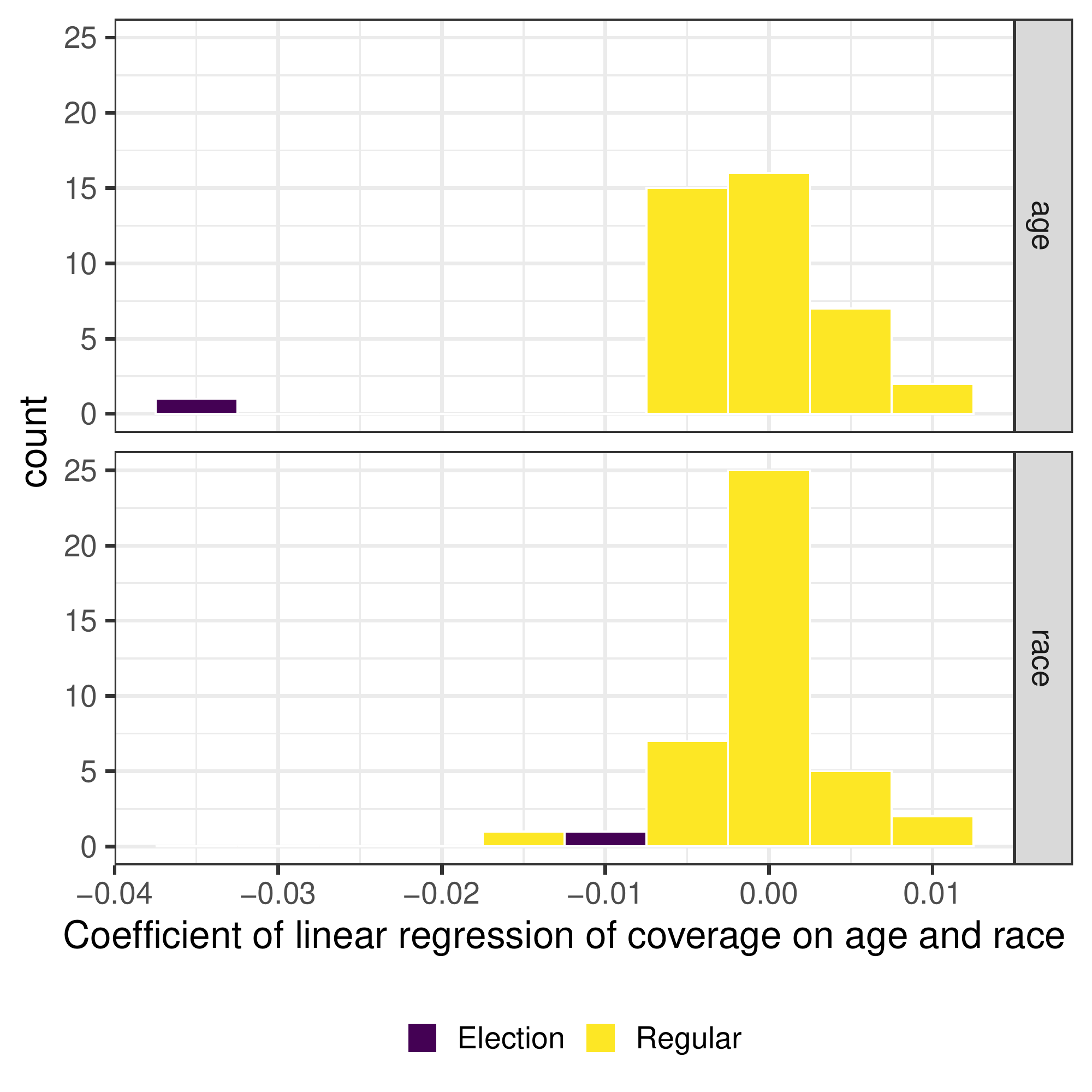}
    \caption{Placebo distribution of coefficients of the linear regression of marginal SafeGraph coverage on election-day age and race demographics. 
     Under the empirical placebo distribution, the election-day's negative coefficients for age and (top panel) and race (bottom panel) are very unlikely ($p$-value $< 0.05$).
     This suggests that SafeGraph data has disparate coverage by age and race. Regressions computed for 40 weekdays of October and November 2018.}
    \label{fig:placebo_coef}
\end{figure}

Figure~\ref{fig:placebo_cor} shows that the negative election-day rank correlation between coverage and voter demographics is significantly outside the placebo distribution for non-election days (empirical one-sided $p$-values are $\approx$0.024 for both age and race, respectively).
For our joint analysis of disparate coverage (Alg.~\ref{alg:joint placebo inference disparate coverage}), we find that the negative coefficients for age and race are statistically outside the placebo distribution (See Fig.~\ref{fig:placebo_coef}; empirical one-sided $p$-values are $\approx$0.024 and 0.049 for age and race respectively).\footnote{In Appendix~\ref{app: assumptions required for measurement validity}, we present similar placebo results for measurement validity.}
Our findings are robust to time-invariant confounding.
Assuming no selection bias (Def.~\ref{assumption:selection}) or time-varying confounding (Def.~\ref{definition: time-varying confounding}), we can conclude that SafeGraph has disparate coverage by age and race, two demographic risk factors for COVID-19.

\subsection{Policy implications}
\label{sec:policy}
We now examine the policy implications of disparate coverage in light of the widespread adoption of SafeGraph data in COVID-19 response. In particular, we show how disparate coverage may lead to under-allocation of important health resources to vulnerable populations. For instance, suppose the policy decision at hand is where to locate mobile pop-up COVID-19 testing sites, and suppose the aim is to place these sites in the most trafficked areas to encourage asymptomatic individuals to get tested. 
One approach could use SafeGraph traffic estimates to rank order POIs. 
How would this ordering compare to the optimal ordering by ground truth traffic?
Using voter turnout as an approximation to ground truth traffic, we perform linear regression of the rank of voter turnout against rank according to SafeGraph marginal traffic as well as age and race: $r(V) \sim r(S-Z) + A + R$. Table~\ref{fig:rank_regression} presents results of this rank regression (where rank is in descending order), confirming that the SafeGraph rank is significantly correlated with ground truth rank.  
But the large coefficient on age indicates that each percentage point increase in voters over 65 is associated with a 5 point drop in rank relative to the optimal ranking. 
Similarly, the coefficient on race indicates that a 1.5 point increase in percent non-white is associated with a one point drop in rank relative to the optimal ranking. 
This demonstrates that ranking by SafeGraph traffic may disproportionately harm older and minority populations by, for instance, failing to locate pop-up testing sites where needed the most.  \nocite{hlavac2018stargazer}

\begin{table}[!htbp] \centering 
 \caption{To evaluate a potential rank-based policy allocation, we compare the rank of voter turnout against rank by SafeGraph traffic, controlling for age and race in a linear regression. Although SafeGraph rank is correlated with the optimal rank by voter turnout, the coefficients on age and race indicate that each demographic percentage point increase is associated with a 5-point and nearly 1-point drop in rank for age and race, respectively. This indicates that significant adjustments based on demographic composition should be made to a SafeGraph ranking. Failure to do so may direct resources away from older and more minority populations.}
    \label{fig:rank_regression} 
\begin{tabular}{@{\extracolsep{5pt}}lc} 
\\[-1.8ex]\hline 
\hline \\[-1.8ex] 
 & \multicolumn{1}{c}{\textit{Dependent variable:}} \\ 
\cline{2-2} 
\\[-1.8ex] & Voter turnout rank \\ 
\hline \\[-1.8ex] 
 SafeGraph rank  & 0.317$^{***}$ \\ 
  & (0.040) \\ 
  & \\ 
 \% over 65 & 4.716$^{***}$ \\ 
  & (0.748) \\ 
  & \\ 
 \% non-white & 0.681$^{**}$ \\ 
  & (0.295) \\ 
  & \\ 
 Constant & 40.278 \\ 
  & (24.830) \\ 
  & \\ 
\hline \\[-1.8ex] 
Observations & 558 \\ 
R$^{2}$ & 0.203 \\ 
Adjusted R$^{2}$ & 0.199 \\ 
Residual Std. Error & 144.264 (df = 554) \\ 
F Statistic & 47.027$^{***}$ (df = 3; 554) \\ 
\hline 
\hline \\[-1.8ex] 
\textit{Note:}  & \multicolumn{1}{r}{$^{*}$p$<$0.1; $^{**}$p$<$0.05; $^{***}$p$<$0.01} \\ 
\end{tabular} 
\end{table} 

We also consider the implications of using SafeGraph to inform proportional resource allocation decisions, such as the provision of masks.
We compare the allocation based on SafeGraph traffic to the allocation based on voter turnout data.
Table~\ref{table_allocation} presents results for polling locations binned into four age-race groups by partitioning at the median proportion over 65 and median proportion non-white.
Each cell presents the proportion of resources that would be allocated to that age-race bin, demonstrating that strict reliance on SafeGraph would under-allocate resources by 37\% 
to the oldest/most non-white category ($p$-value < 0.05) and over-allocate resources by 33\% to the youngest/whitest category ($p$-value < 0.05).

\begin{table}[ht]
\centering
\begin{tabular}{rccccc}
  \hline
 & SafeGraph & Optimal  & Percent\\ 
 & allocation & allocation & difference \\ 
  \hline
young white & 0.33  & 0.25  & +33\%\\ 
 &  (0.03) & (0.02) \\ 
  young non-white & 0.33 & 0.35 & -5\% \\ 
 & (0.03) & (0.03) \\ 
 older white & 0.19 & 0.18  & +9\%\\ 
  & (0.02) & (0.01) \\ 
 older non-white & 0.13 & 0.21 & -37\%\\ 
  & (0.02) & (0.02) \\ 
   \hline
\end{tabular}
\caption{Allocation of resources for age-race groups by SafeGraph versus by true voter counts, with standard errors in parentheses. The SafeGraph allocation redirects over one-third of the optimal allocation from the oldest, most non-white group to the youngest, whitest group  ($p$-value < 0.05). } 
\label{table_allocation}
\end{table}

The clear policy implication here is that while SafeGraph information may aid in a policy decision, auxiliary information (including prior knowledge) should likely be combined to make final resource allocation decisions.

\section{Discussion}
\label{sec:discuss}
We have provided the first independent audit of demographic bias of a smartphone-based mobility dataset that has been widely used in the policy response to COVID-19. 
Our audit indicates that the data underrepresents two high risk groups: older and more non-white populations.
Our results suggest that policies made without adjustment for this sampling bias may disproportionately harm these high risk groups. 
However, we note a limitation to our analysis.
Because SafeGraph information is aggregated for privacy reasons, we are not able to test coverage at the individual level. 
To avoid a potential ecological fallacy, our results should be interpreted as a statement about \emph{POIs} rather than \emph{individuals}.
That is, POIs frequented by older (or minority) visitors have lower coverage than POIs frequented by younger (or whiter) populations.
Of course, policy decisions are typically made at some level of aggregation, so the demographic bias we document at this level remains relevant for those decisions.

A key future research question is how to use the results of this audit to improve policy decisions.
We suggest a few possible future directions.
A bias correction approach would construct weights to adjust estimates based on race and age.
Such an approach crucially requires knowledge about demographic composition.
In policy settings where such information is not readily available, it may be fruitful to investigate whether mobility data companies like SafeGraph can provide normalized visit counts based on the estimated demographic profile of the smartphone user. This could offer a significant improvement over current normalization approaches which, per SafeGraph's recommendation, 
 use census block group (CBG)-based normalization factors \citep{squire2019measuring}.  
While this bias correction might help to estimate population parameters (e.g., percentage of CBG population not abiding by social distancing), it is unlikely to capture the kind of demographic interaction effects we document here. 
Much more work should be done to study disparate coverage and ideally provide, for instance, a weighing correction to the normalization factors that properly accounts for the demographic disparities documented in this audit.
 
Another possible solution is increased transparency. 
Researchers do not know details about the source of SafeGraph's mobility data, namely which mobile apps feed into the SafeGraph ecosystem.
Access to such information may make the bias correction approach more tractable.
If, for instance, researchers could identify that a data point emanates from Snapchat, then they could use what is known about the Snapchat user base to make adjustments.   
Given its increasing importance for policy, SafeGraph should consider disclosing more details about which apps feed into their ecosystem.

\section{Conclusion}
\label{sec:conclude}

Mobility data based on smartphones has been rapidly adopted in the COVID-19 response.  As \cite{grantz2020use} note, one of the most profound challenges arising with such rapid adoption has been the need to assess the potential for demographic bias ``when there is no clear gold standard against which to validate mobile phone data.'' Our paper illustrates one potential path forward, by linking smartphone-based data to high-fidelity ground truth administrative data.  Voter turnout records, which record at the individual level whether a registered voter traveled to a polling location on a specific day and describe the voter's demographic information, enable us to develop a straightforward audit test for disparate coverage.  We find that coverage is notably skewed along race and age demographics, both of which are significant risk factors for COVID-19 related mortality.  Failure to address such disparities risks policy distortions based on mobility data that could exacerbate serious existing inequities in the health care response to the pandemic.

\begin{acks}
We thank SafeGraph for making their data available, answering our many questions, and providing helpful feedback. 
We are grateful to Stanford's Institute for Human-Centered Artificial Intelligence, the Stanford RISE initiative, the K\&L Gates Presidential Fellowship, and the National Science Foundation for supporting this research.
This material is based upon work supported by the the National Science Foundation  Graduate
Research Fellowship Program under Grant No. DGE1745016. Any opinions,
findings, and conclusions or recommendations expressed in this material are those of the
author(s) and do not necessarily reflect the views of the National Science Foundation.
We gratefully acknowledge Mark Krass for first suggesting voter turnout data. We thank Angie Peng, Rayid Ghani, and Dave Choi for providing helpful feedback.
\end{acks}

\clearpage

\bibliographystyle{ACM-Reference-Format}
\bibliography{ref}

\clearpage

\appendix
\section*{Appendix}
\section{Data}
\label{app:data}

\subsection{Mobility Data}
Our mobility data comes from SafeGraph via its COVID-19 Data Consortium.  Specifically, we rely on the SafeGraph Patterns data, which provides daily foot traffic estimates to individual POIs, and the Core Places data, which contains basic location information for POIs.

\subsection{Election Data}
Our election data comes from certified turnout results of the 2018 North Carolina general election, as collected by L2. For each registered voter, L2 provides demographic data, such as name, age, ethnicity, and voting district/precinct, as well as their voter history.

\begin{figure}[ht]
    \centering
    \includegraphics[scale=0.3]{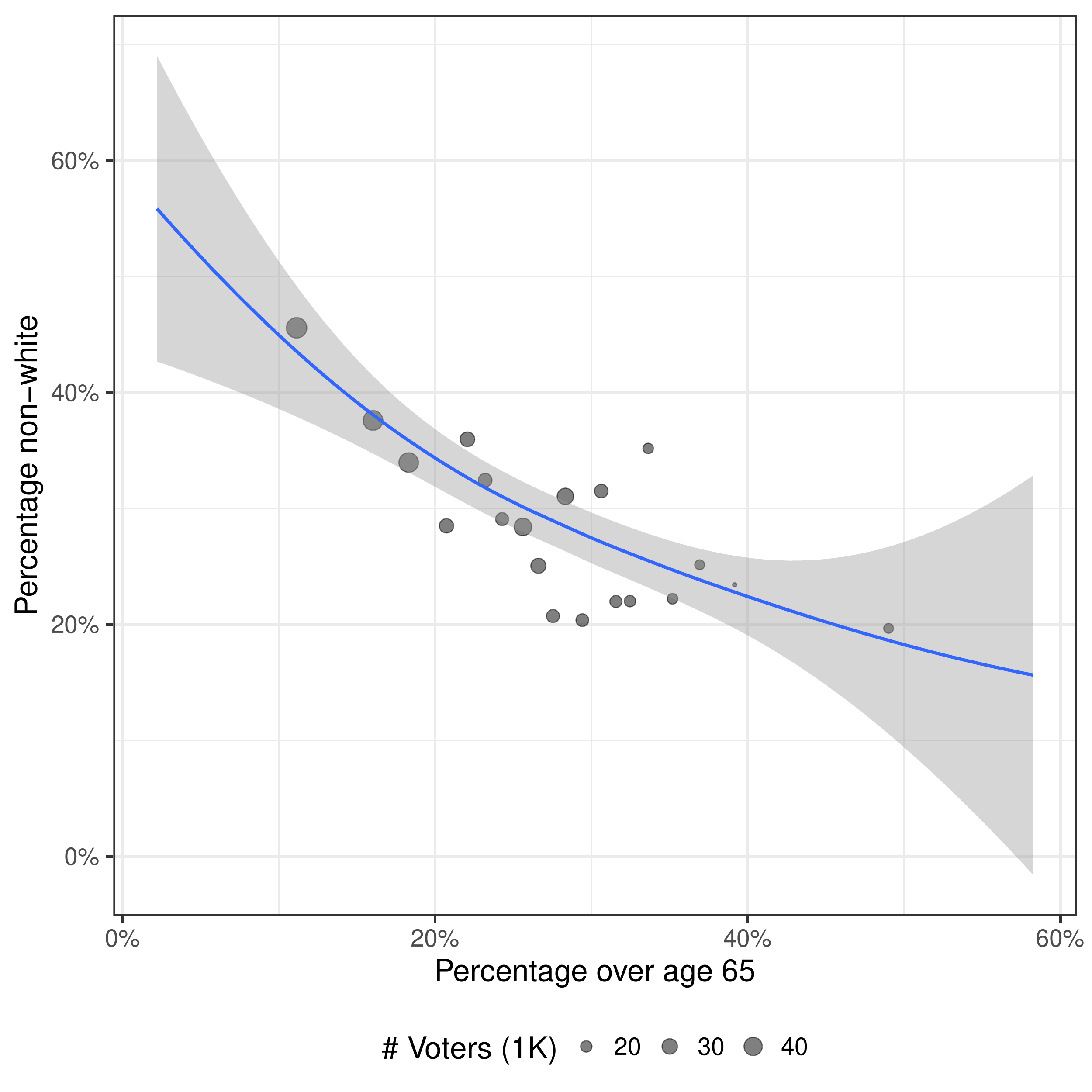}
    \caption{Non-white voters are more likely to be young}
    \label{fig:age_race}
\end{figure}

We provide some additional descriptive information about the data here. First, Figure~\ref{fig:age_race} shows the correlation between age and race across polling locations.  This illustrates the importance of jointly interpreting how coverage varies by age and race. 

\begin{figure}[ht]
    \centering
    \includegraphics[scale=0.3]{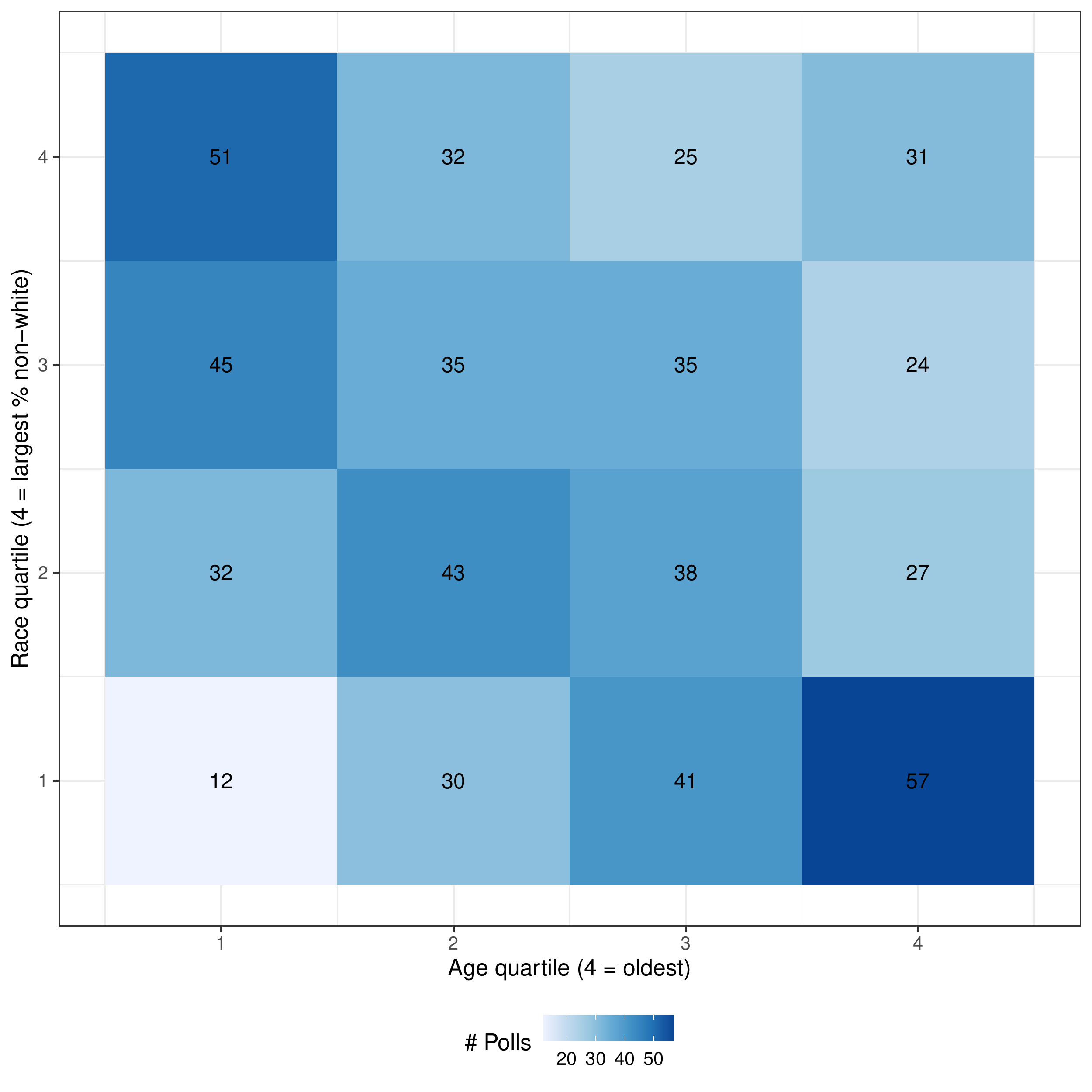}
    \includegraphics[scale=0.3]{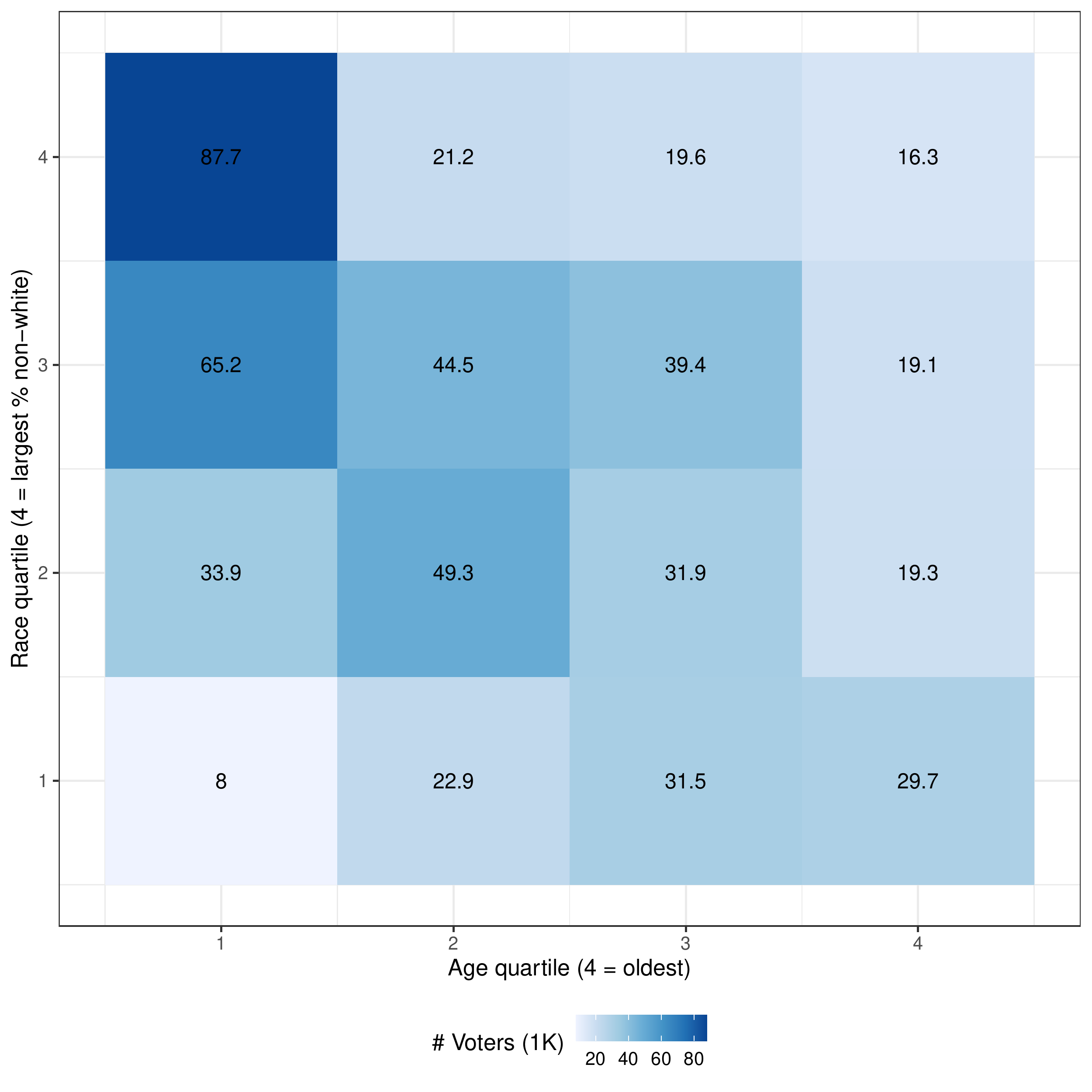}
    \caption{Joint distribution polling locations and voters by age quartiles ($x$-axis) and race quartiles ($y$-axis).}
    \label{fig:count_map}
\end{figure}

Second, the top panel of Figure~\ref{fig:count_map} illustrates the density of locations by age quartile on the $x$-axis and race quartile on the $y$-axis.  The two modal polling locations are for locations with white elderly populations and non-white young populations.  The bottom panel displays the total number of voters (in units of 1000) in these cells, showing that young, high-minority cells represent a particularly large number of voters. 

\subsection{Poll Location Data}
Our polling location and precinct data for North Carolina for Election Day 2018 was acquired from the North Carolina Secretary of State. This dataset contains the street address for each polling place, including location name, county, house number, street name, city, state, and zip code, as well as the precinct associated with the polling location.


\section{Details on data cleaning and merging}
\label{app:preprocess}
This study required that we merge the points-of-interest (POIs) as defined by SafeGraph with the polling locations in North Carolina in 2018. To do so, we used SafeGraph’s Match Service\footnote{See \url{https://docs.safegraph.com/docs/matching-service-overview}.}, which takes in a POI dataset and using an undisclosed algorithm, matches it with its list of all POIs, appending at least one SafeGraph ID for all matched POIs.\footnote{In some cases, SafeGraph appends multiple candidate POI IDs to a given input location. We removed such cases since we could not confidently ascertain which candidate was the correct match.} The service utilizes a variety of basic information\footnote{See \url{https://docs.safegraph.com/v4.0/docs/places-schema}.} to determine matches; of these, we provided the location name (or polling place name), street address, city, state, and postal code for all polling locations in North Carolina in 2018. The match rate, i.e. the percentage of input polling locations SafeGraph could match with one of its POIs, was 77.6\%.

The polling location dataset, now having SafeGraph IDs for each matched location, was then joined with the SafeGraph Places dataset, which contains basic information like location name and address for the POI, for comparison between the matched POI and the polling location. The SafeGraph matching algorithm was at times too lenient, matching locations near each other but with different names or matching locations with different street addresses. To remedy this, we ran the dataset through a postprocessing script which removed matches where the street addresses (street number and street names) differed by three or more tokens to account for false positives.\footnote{For example, "1100 Auto Center Circle" would be considered to have four tokens.} This resulted in a match rate of 47.7\%. We then filtered out POIs where SafeGraph returned multiple candidate matches since we could not be confident the first match was the correct match. This resulted in a match rate of 42.4\%.

Next, we mapped voters from the L2 voter file to the appropriate polling location with SafeGraph ID. The L2 voter file contains the precinct for each voter, and the polling location data associates each precinct with a polling location, so by mapping voter to precinct and precinct to polling location (and SafeGraph ID) we could fetch the polling location for each voter for which there was a match with a SafeGraph POI. We observed differences in how the polling data and L2 named the same precinct. For instance, one source may using preceding zeros "0003" whereas another may not "3" or one source may use  "WASHINGTON WARD 1" whereas the second uses "WASHINGTON 1". We manually resolved some of these discrepancies, but we were unable to resolve all and therefore had to drop the corresponding poll location. After this filtering, we obtained a poll location match rate of 30.1\%.

Our final preprocessing step removed polling locations at elementary and secondary schools because we found there was too much variation in traffic to reliably impute the non-voter traffic. For instance, we find 169 instances in which the difference in POI traffic between two adjacent days is over 100, and all 169 such instances pertain to schools. As shown in the below table, a linear regression model estimating traffic for a given day using the traffic on adjacent days as features (which is later used in section ~\ref{app:bandwidth}) has higher RMSE when used to predict traffic for school POIs, even when trained on school POIs, compared to non-school POIs. This shows that traffic on adjacent days is less effective for imputing traffic on a given day for schools and that imputing traffic for schools would require a different approach. Filtering out polling locations at schools resulted in a final sample of 20.8\% of the polling locations in North Carolina's 2018 general election.

\begin{center}
\begin{tabular}{llc}
 \hline
 Prediction target & Training data & RMSE\\
 \hline
 Schools & Schools & 14.686\\
  & No schools & 15.044\\
 All but schools & Schools & 5.508\\
     & No schools & 5.375\\
 \hline
\end{tabular}
\end{center}

\subsection{Challenges for scalability}
While a key virtue of our approach is bringing in auxiliary ground truth data, the drawback is that this approach is not conducive to iterative audits over time (or geography) because of scalability challenges.
Voter locations change with every election and there is no national database that collects voter location information over time. Creating the crosswalks between (a) SafeGraph POIs and voter locations and (b) voter locations and precincts in voter turnout files is a heavily manual process that differs for each jurisdiction, given the decentralized nature of election administration. 

\section{Assumptions required for measurement validity} \label{app: assumptions required for measurement validity}
The assumptions required for our measurement validity analysis are much weaker than those discussed and evaluated in the main paper, but we provide the results here for completeness.
To identify the relationship between ground truth visits and SafeGraph traffic, we need the following to hold:
\begin{assumption}[No induced confounding (measurement validity)] \label{assumption:confounding measurement}
  The estimation procedure does not induce a confounding factor that affects both the estimates of ground truth visits $T$ and the estimated marginal SafeGraph traffic $S - Z$.
\end{assumption}
\begin{assumption}[No selection bias (measurement validity)]
\label{assumption:selection measurement}
  The selection is not based on an interaction between factors that affect ground truth visits $T$ and the estimated marginal SafeGraph traffic $S - Z$.
\end{assumption}

As we do above for disparate coverage, we can partially test Assumption~\ref{assumption:confounding measurement} using placebo inference (see next section). While we can test for time-invariant confounding, we cannot test for time-varying confounding. 
Nonetheless it is difficult to postulate a reasonable mechanism for time-varying confounding in our measurement analysis.
Assumption~\ref{assumption:selection measurement} would be violated if SafeGraph coverage is better for polling location POIs versus non-polling location POIs. 

\subsection{Robustness test for time-invariant confounding}
\begin{figure}
    \centering
    \includegraphics[scale = 0.3]{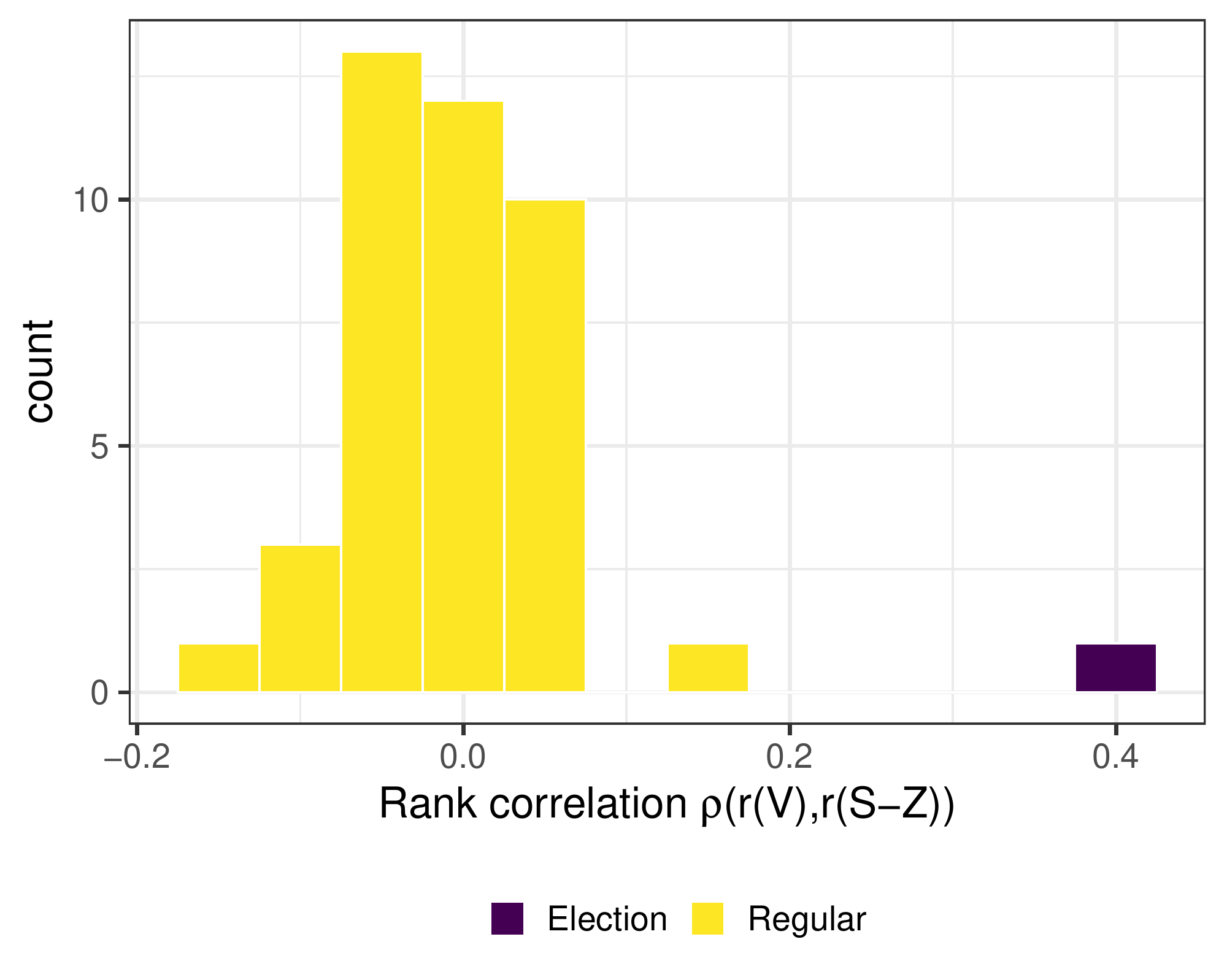}
    \caption{Placebo distribution of rank correlation between voters and marginal SafeGraph traffic. As expected, voter turnout is positively correlated with SafeGraph marginal traffic only on election day (empirical $p$-value = 0.024).}
    \label{fig:placebo_cor_count}
\end{figure}

To allow for time-invariant confounding in our estimation of the correlation between ground truth visits and SafeGraph visits, we consider the rank correlation between voter turnout and SafeGraph marginal traffic on non-election days (Alg.~\ref{alg:placebo inference measurement} provides details).
We would not expect to find a non-zero such correlation, and indeed Figure~\ref{fig:placebo_cor_count} shows that the positive correlation on election day is significantly outside the distribution for placebo days (empirical one-sided $p$-value = 0.024).

\begin{algorithm}[htbp!]
\SetAlgoLined
\KwIn{Voter data $(V^{j*})$
SafeGraph data $\{(S^{j}, Z^{j})\}_{j=1}^{n}$
}
\KwResult{$p$-value for the election-day correlation under the placebo distribution}
\For{$j = 1,2, \hdots n$}{ 
Compute $\rho_j = \cor(r(S^{j} - Z^{j}), r(V^{j*}))$.
}
\Return{$p = \frac{1}{n} \displaystyle \sum_{j =1}^n\mathbb{I}\{(\rho_j \geq \rho_{j^{*}})\}$}
\caption{Assessing measurement validity  (Def.~\ref{definition: robust disparate coverage})}
\label{alg:placebo inference measurement}
\end{algorithm}

\section{Regression Imputation of Non-Voter Traffic}
\label{app:bandwidth}
In this section, we consider various approaches for estimating non-voter traffic and we confirm that our disparate coverage results hold under these approaches. At a high-level, we estimate non-voter traffic on election day using the number of visits on adjacent weekdays. We first consider the mean imputation approach and then proceed to a linear regression approach. For both approaches, we identify the optimal number of adjacent weekdays by evaluating how well we estimate the amount of traffic on non-election day Tuesdays from January 2018 to April 2020.

For the mean imputation approach, we look at the $X$ adjacent weekdays before and after a given Tuesday and use the average of the traffic on all those weekdays to estimate the traffic on Tuesday. That is, for $X = 2$, we calculate the estimate as the average of the traffic on the Friday and Monday before and Wednesday and Thursday after a given Tuesday. We performed this calculation for all North Carolina polling locations and all Tuesdays, excluding Election Days and the first and last Tuesdays from January 2018 to April 2020, with traffic data available from the SafeGraph Patterns data, which gave us 147,613 data points. We tested $X \in [1,4]$ as this considers all weekdays up to the next or previous Tuesday. The following are the evaluation metrics for this approach:
\begin{center}
\begin{tabular}{ |c||c|c|c|  }
 \hline
 $X$ & RMSE & R2 & MAE\\
 \hline
 1 & 10.85 & 0.870 & 4.05\\
 2 & 10.37 & 0.881 & 3.91\\
 3 & 10.61 & 0.875 & 3.95\\
 4 & 10.75 & 0.874 & 4.00\\
 \hline
\end{tabular}
\end{center}

\begin{figure}[h!]
     \centering
     \includegraphics[scale=.26]{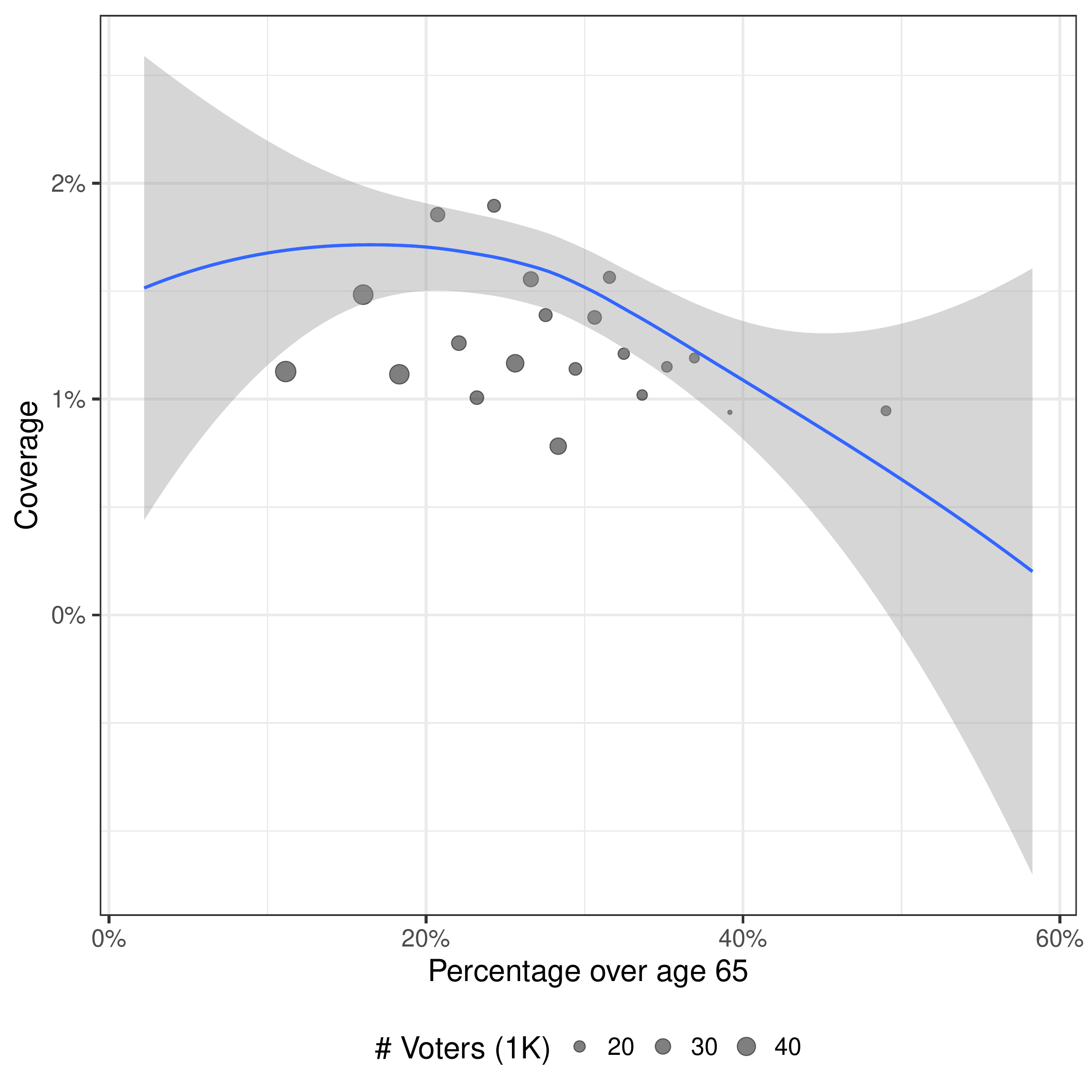}
    \includegraphics[scale=0.26]{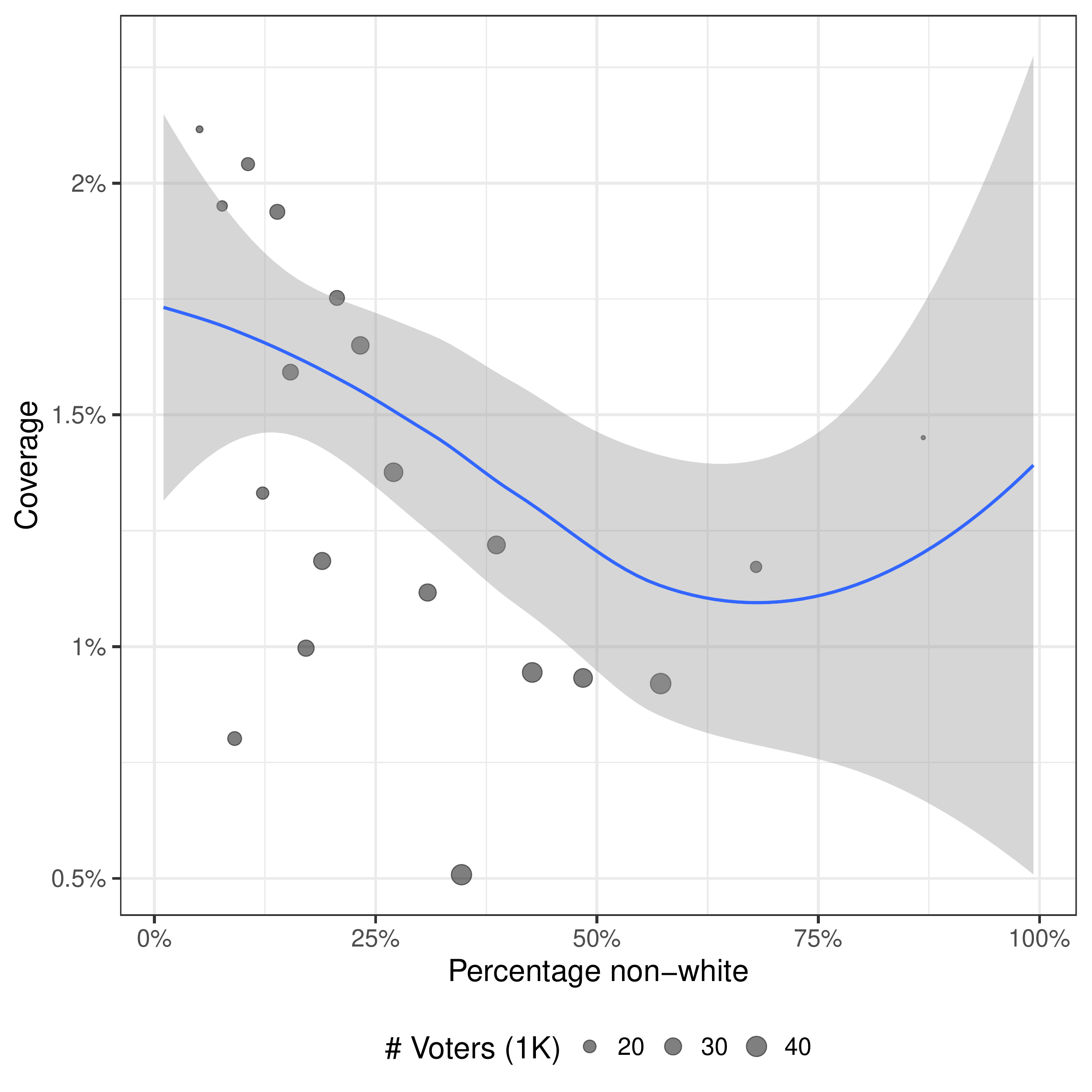}
    \caption{Estimated SafeGraph coverage rates against age and race for North Carolina 2018 general election for ventiles of poll location by age (top) and race (bottom) when using the optimal ($X=2$) adjacent days to impute non-voter traffic.}
    \label{fig:adj_2_correlations}
 \end{figure}

\begin{figure}[t]
    \centering
    \includegraphics[scale=0.3]{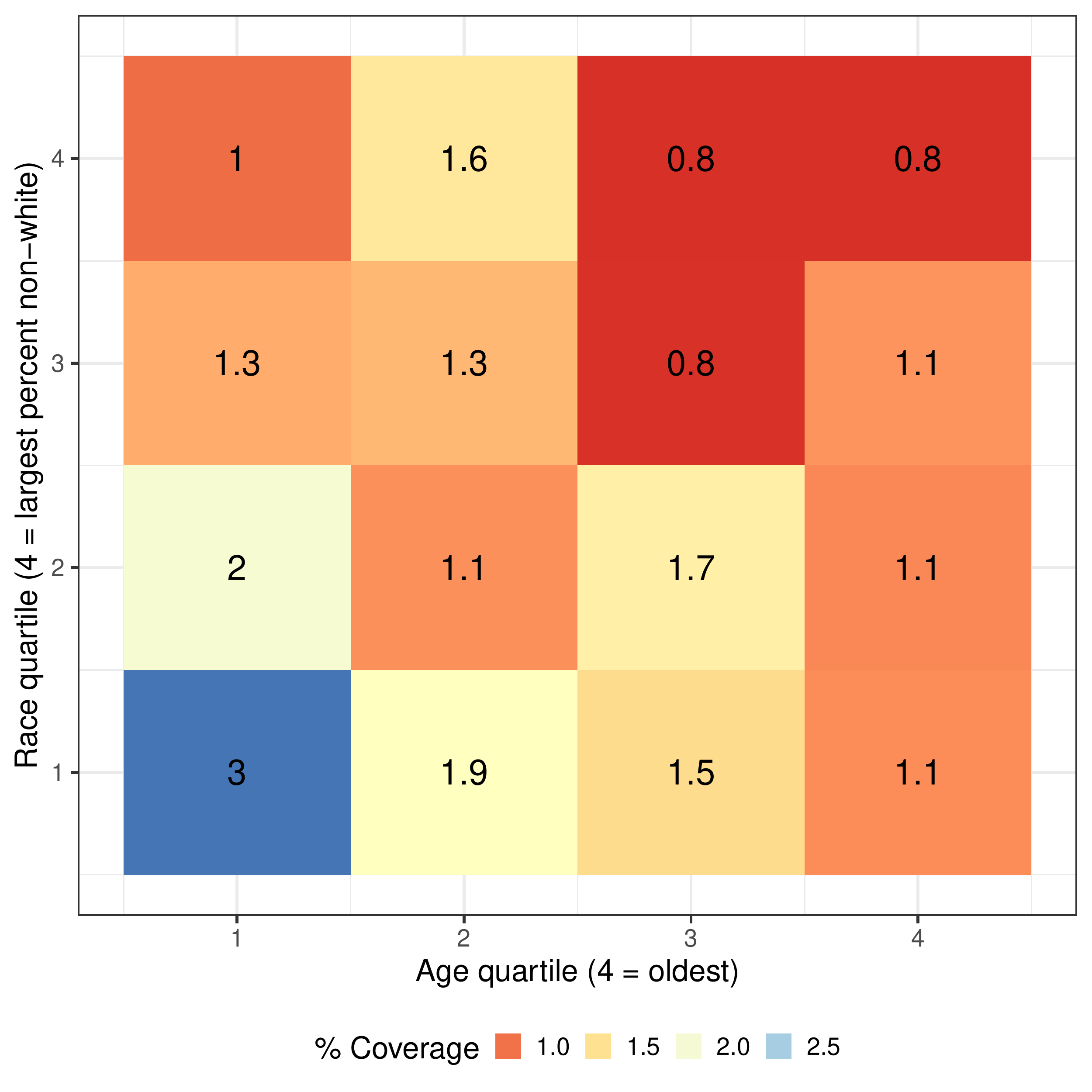}
    \includegraphics[scale= 0.3]{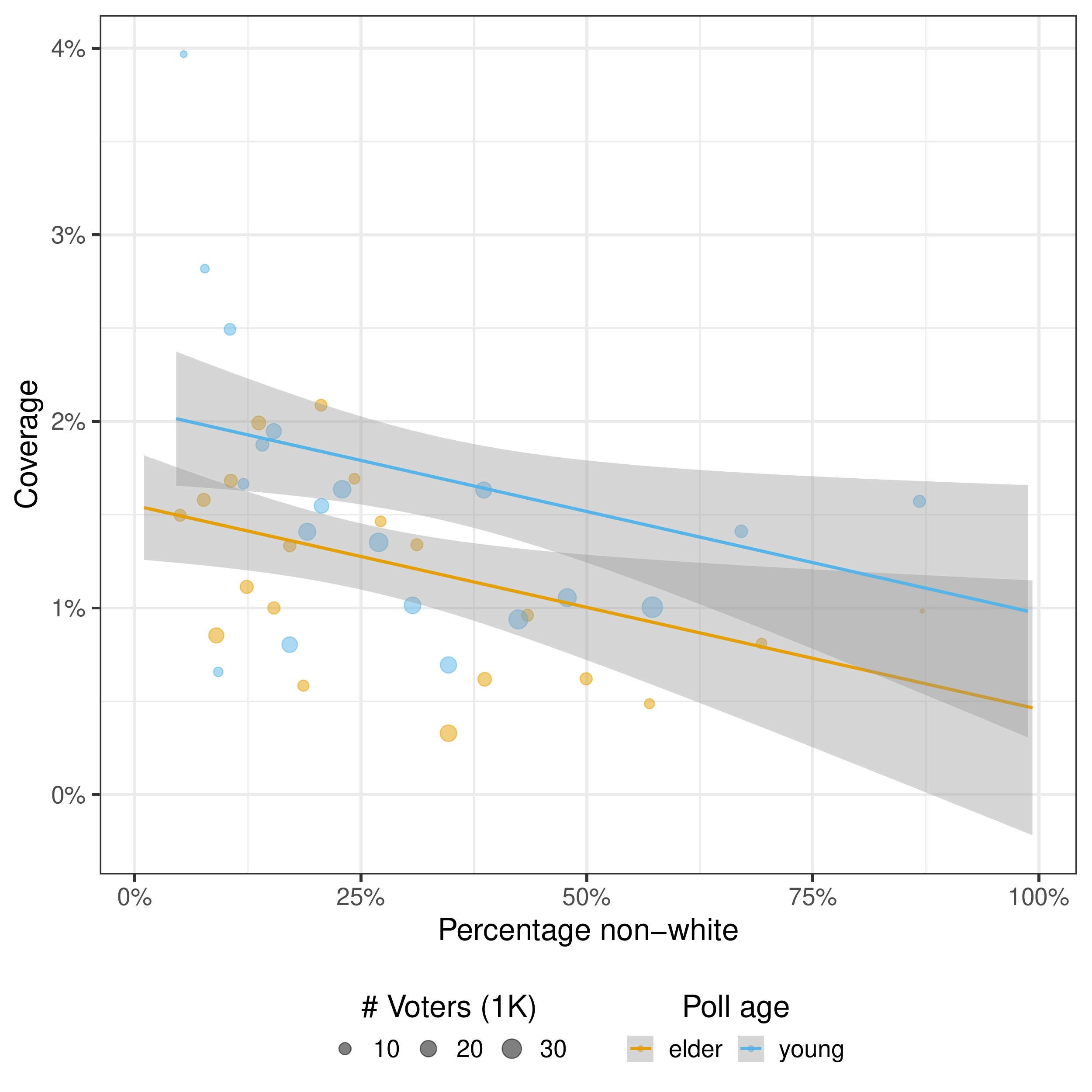}
    \caption{Intersectional coverage effects by race and age when using the optimal ($X=2$) adjacent days to impute non-voter traffic. The top panel presents the coverage rate by quartiles of age on the $x$-axis and race on the $y$-axis.  The bottom panel plots the coverage rate on the $y$-axis against percentage of non-white voters at the polling location on the $x$-axis for older polling locations (yellow) versus younger polling locations (blue) for ventiles of poll location by race. (Lines display linear smoothing of the individual poll locations.) Coverage is lowest among older minority populations and highest among younger whiter populations. }
    \label{fig:adj_2_heat_map}
\end{figure}

Averaging the traffic on the two weekdays before and after a given Tuesday performs best by all three evaluation metrics. Repeating our analysis using the two weekdays before and after election day to estimate $Z$ yields comparable results to those presented in the main paper.
The rank correlation tests for age and race respectively yield $\cor\big( r(C(S-Z,V), r(D) \big) = -0.13$ and $-0.11$ ($p$-value $<0.01$).
 Fig.~\ref{fig:adj_2_correlations}-\ref{fig:adj_2_heat_map} provide the analogous results to Fig.~\ref{fig:age_rate}-\ref{fig:heat_map} from the main paper.

In the second approach, we used the traffic on adjacent weekdays as features for a linear regression model, to account for the possibility that traffic on certain weekdays may be more impactful in calculating an accurate estimate. With the same dataset as the one described for the first approach we used 10-fold cross validation with 3 repeats, with the following results:
\begin{center}
\begin{tabular}{ |c||c|c|c|  }
 \hline
 $X$ & RMSE & R2 & MAE\\
 \hline
 1 & 10.65 & 0.872 & 4.16\\
 2 & 9.98 & 0.888 & 3.89\\
 3 & 9.88 & 0.890 & 3.83\\
 4 & 9.76 & 0.893 & 3.76\\
 \hline
\end{tabular}
\end{center}
The linear model using traffic from the four adjacent weekdays before and after the given Tuesday performed the best across both approaches, so we used this model to estimate non-voter traffic on Election Day.

We used the model to predict the number of non-voter visits to each of 558 polling location POIs on Election Day (November 6, 2018). 10 of the POIs did not have Patterns visit count data for 4 weekdays before Election Day (October 31, 2018), so we imputed the traffic to be the traffic 3 weekdays before (November 1, 2018) for those POIs. Using this imputation scheme, the rank correlation tests for age and race respectively yield $\cor\big( r(C(S-Z,V), r(D) \big) = -0.13$ ($p$-value $<0.01$) and $-0.10$ ($p$-value $<0.05$).

The model was also used to impute the traffic at poll location POIs on 40 weekdays between October 1, 2018 and November 30. These predictions were then used to repeat the data analyses on SafeGraph coverage as an additional robustness check, producing similar results to original analysis that relied on mean imputation to adjust for non-voter traffic. 

\section{Additional Results}
\label{app:additional_results}

Table~\ref{tb:interaction_reg} presents regression results for fitting coverage on election-day as a linear function of the percent of the voting population over 65 ($A^{j*}$), the percentage of the population that is non-white ($R^{j*}$), and their interaction: $$C(S^{j*} - Z^{j*},V^{j*}) = \beta_0 +  \beta_1 A^{j*},  + \beta_2 R^{j*} + \beta_3 A^{j*} *R^{j*}$$
The first column shows that the percent of the voting population over 65 is negatively associated with coverage.  The second column shows that controlling for age, an increase in the percentage of the population that is non-white is associated with a decrease in the coverage rate.  The third column fits interaction terms. 


\begin{table*}[h] \centering 
  \caption{} 
  \label{tb:interaction_reg}  
\begin{tabular}{@{\extracolsep{5pt}}lccc} 
\\[-1.8ex]\hline 
\hline \\[-1.8ex] 
 & \multicolumn{3}{c}{\textit{Dependent variable:}} \\ 
\cline{2-4} 
\\[-1.8ex] & \multicolumn{3}{c}{rate} \\ 
\\[-1.8ex] & (1) & (2) & (3)\\ 
\hline \\[-1.8ex] 
 \% over 65 & $-$0.028$^{***}$ & $-$0.035$^{***}$ & $-$0.024$^{*}$ \\ 
  & (0.008) & (0.008) & (0.014) \\ 
  & & & \\ 
  \% non-white &  & $-$0.012$^{***}$ & $-$0.002 \\ 
  &  & (0.003) & (0.011) \\ 
  & & & \\ 
 \% non-white $\times$ \% over 65 &  &  & $-$0.0004 \\ 
  &  &  & (0.0004) \\ 
  & & & \\ 
 Constant & 2.269$^{***}$ & 2.834$^{***}$ & 2.529$^{***}$ \\ 
  & (0.242) & (0.287) & (0.418) \\ 
  & & & \\ 
\hline \\[-1.8ex] 
Observations & 558 & 558 & 558 \\ 
R$^{2}$ & 0.020 & 0.042 & 0.044 \\ 
Adjusted R$^{2}$ & 0.018 & 0.038 & 0.038 \\ 
Residual Std. Error & 1.707 (df = 556) & 1.689 (df = 555) & 1.689 (df = 554) \\ 
F Statistic & 11.207$^{***}$ (df = 1; 556) & 12.102$^{***}$ (df = 2; 555) & 8.403$^{***}$ (df = 3; 554) \\ 
\hline 
\hline \\[-1.8ex] 
\textit{Note:}  & \multicolumn{3}{r}{$^{*}$p$<$0.1; $^{**}$p$<$0.05; $^{***}$p$<$0.01} \\ 
\end{tabular} 
\caption{Linear regression models of coverage rate by demographic attributes of polling locations.}
\end{table*} 
\end{document}